\documentclass[%twocolumn,%showpacs,preprintnumbers,
amsmath,%amssymb,aps,
prd,nofootinbib,floatfix,11pt,%preprint,11pt
]{revtex4}

\newcommand\beq{\begin{eqnarray}}
\newcommand\eeq{\end{eqnarray}}

\def\lsim{\mathrel{\rlap{\lower4pt\hbox{$\sim$}}
    \raise1pt\hbox{$<$}}}                % less than or approx. symbol
\def\gsim{\mathrel{\rlap{\lower4pt\hbox{$\sim$}}
    \raise1pt\hbox{$>$}}}            

\def\dilog{{\rm Li}_2}
\def\epsUV{\epsilon_{\rm UV}}
\def\epsIR{\epsilon_{\rm IR}}
\def\MSbar{$\overline{\rm MS}$ }
\def\lntwo{\ln(2)}
\allowdisplaybreaks
\usepackage{graphicx}% Include figure files
\usepackage{setspace}
\usepackage{axodraw}

\begin{document}
\renewcommand{\theequation}{\arabic{section}.\arabic{equation}}

\title{\Large%
Radiative corrections to stoponium annihilation decays}
\author{Stephen P. Martin$^{1,2}$ and James E. Younkin$^1$}
\affiliation{$^1$Department of Physics, Northern Illinois University, 
DeKalb IL 60115}
\affiliation{$^2$Fermi National Accelerator Laboratory, P.O. Box 500, 
Batavia IL 60510}

%\date{\today}

\begin{abstract}
The lighter top squark in supersymmetry can live long enough to form 
hadronic bound states if it has no kinematically allowed two-body decays 
that conserve flavor. In this case, scalar stoponium may be observable 
through its diphoton decay mode at the CERN Large Hadron Collider, 
enabling a uniquely precise measurement of the top-squark mass. The 
viability of the signal depends crucially on the branching ratio to 
diphotons. We compute the next-to-leading order QCD radiative corrections 
to stoponium annihilation decays to hadrons, photons, and Higgs scalar 
bosons. We find that the effect of these corrections is to significantly 
decrease the 
predicted branching ratio to the important diphoton channel. We also find 
a greatly improved renormalization-scale dependence of the 
diphoton branching ratio prediction.
\end{abstract}

\maketitle
\tableofcontents
\baselineskip=14.9pt

\newpage
\setcounter{footnote}{1}
\setcounter{page}{1}
\setcounter{figure}{0}
\setcounter{table}{0}

%%%%%%%%%%%%%%%%%%%%%%%%%%%%%%%%%%%%%%%%%%%%%%%%%%%%%%%%%%%%%%%%%%%%%%%%%%%%%%%%
%%%%%%%%%%%%%%%%%%%%%%%%%%%%%%%%%%%%%%%%%%%%%%%%%%%%%%%%%%%%%%%%%%%%%%%%%%%%%%%%
\section{Introduction\label{sec:intro}}
\setcounter{footnote}{1}
\setcounter{equation}{0}

The Minimal Supersymmetric Standard Model (MSSM) \cite{reviews} with 
conserved R-parity contains a stable lightest supersymmetric particle 
(LSP). If the LSP is neutral, it could be the cold dark matter required 
by the standard cosmology. The collider signatures of the MSSM generally 
involve missing energy carried away by two LSPs produced in each event. 
Unfortunately, this suggests that there will be no true kinematic mass 
peaks whose reconstruction would determine superpartner masses. In 
favorable models, it is possible to obtain precision measurements of 
superpartner mass differences and other combinations of masses at hadron 
colliders by finding kinematic edges from decays. However, the overall 
mass scale of the superpartners will be much harder to obtain precisely
at the CERN Large Hadron Collider \cite{ATLASTDR,CMSTDR}.

In models with a relatively small mass difference between the lighter top 
squark $\tilde t_1$ and the neutralino LSP $\tilde N_1$, there is an 
exception that would allow a sharp mass peak. If the lighter top squark 
has no kinematically allowed two-body decays that conserve flavor, then 
it will form hadronic bound states. Among these is stoponium, a 
stop-anti-stop bound state, which can be directly produced at hadron 
colliders through gluon-gluon fusion. The largest production 
cross-section is for the $1S$ ($J^{PC} = 0^{++}$) state, denoted in the 
following by $\eta_{\tilde t}$, but other stoponium states can contribute 
to the signal either by prompt decays to the ground state or direct 
annihilation decays. This state will form if
\beq
m_{\tilde t_1} &<& m_{\tilde N_1} + m_t,
\label{eq:stoplNt}
\\
m_{\tilde t_1} &<& m_{\tilde C_1} + \mbox{5 GeV},
\label{eq:stoplCb}
\eeq
so that the decays
$\tilde t_1 \rightarrow t \tilde N_1$ and
$\tilde t_1 \rightarrow b \tilde C_1$
are both kinematically forbidden. 
(Here $\tilde C_1$ stands for the lighter chargino mass eigenstate.)
These conditions are almost never satisfied in the MSSM parameter
space with the so-called mSUGRA boundary conditions, but they can
easily be satisfied in other motivated models. These include 
``compressed supersymmetry"
models 
\cite{compressed}
in which the predicted thermal relic density of dark matter is in
agreement with that observed by WMAP
and other experiments
\cite{WMAP}-\cite{PDG}, due to the enhanced annihilation
$\tilde N_1 \tilde N_1 \rightarrow t \overline t$ mediated by $t$-channel
exchange of top squarks.\footnote{The inequalities (\ref{eq:stoplNt})
and (\ref{eq:stoplCb}) can also be satisfied in
the stop-neutralino co-annihilation region
\cite{stopcoannihilationone}-\cite{stopcoannihilationthree}
of parameter space, but at the price of a much more extreme fine-tuning
of input parameters.}
Another class of models consists of those that generate
the baryon excess over anti-baryons at  
the electroweak scale \cite{baryo}, \cite{baryoDM}, \cite{baryonew}. 
%% version 2: next line has reference added
In these and other \cite{DiazCruz:2007fc} cases,
$m_{\tilde t_1} - m_{\tilde N_1}$ is necessarily small enough to
guarantee the formation of stoponium.

Stoponium can decay directly by the decays 
of one of the top-squark constituents
through the 3-body process 
$\tilde t_1 \rightarrow W b \tilde N_1$, 
or if that is kinematically forbidden, through the
flavor-violating 2-body process
$\tilde t_1 \rightarrow c \tilde N_1$ and/or the
4-body process $\tilde t_1 \rightarrow f \overline f' b \tilde N_1$.
However, the corresponding partial widths are many orders of magnitude 
smaller
\cite{Hikasa:1987db}-\cite{Hiller:2008wp}
than the binding energy of stoponium, which will be of order a few GeV
\cite{Hagiwara:1990sq}. 
Therefore, $\eta_{\tilde t}$ will decay primarily by annihilation,
including the possible two-body final states
$gg$, $\gamma\gamma$, $h^0h^0$, $W^+W^-$, $ZZ$, $Z\gamma$, $t \overline t$,
and $b \overline b$. Of these, the most promising final state, both
for detectability over backgrounds   
and reconstruction of the mass peak, is $\gamma\gamma$,
as was first pointed out long ago by Drees and Nojiri
\cite{Drees:1993yr,Drees:1993uw}
(See also refs.~\cite{Nappi:1981ft}-\cite{Fabiano:2001cw}
for other work related to stoponium at colliders.)
The diphoton stoponium signals for both compressed supersymmetry and 
supersymmetric
electroweak-scale baryogenesis have recently been studied in 
\cite{Martin:2008sv}.

In much of the parameter space in which stoponium can form,
the $gg$ two-body final state leading to hadronic jets dominates.
If so, then the leading-order
prediction for BR$(\gamma\gamma)$ is nearly model-independent, and
is of order $0.005$. The QCD corrections to the bound state annihilation
decays are quite significant, however, and need to be taken into account
along with corrections to the production cross-section in order to
obtain a realistic estimate of the LHC sensitivity for a given model.   
In this paper, we will calculate the QCD next-to-leading order 
corrections to 
$S$-wave stoponium\footnote{Although we have in mind the top squark, 
our results can also be applied to 
any new strongly interacting fundamental scalar whose width is small
enough that it hadronizes before it decays.}  
decay into the $\gamma\gamma$, $gg$, and $h^0h^0$ final states.
The decay widths into $\gamma\gamma$ and $gg$ are model-independent to
leading order, and one may also argue that they are the most important,
since $gg$ usually dominates the total width, and $\gamma\gamma$ is the 
observable
signal. However, the $h^0h^0$ final state may dominate in some parts 
of parameter space, 
particularly if stoponium is just above the 
threshold for that decay.

We will proceed following the strategy (and some of the notation) used in 
ref.~\cite{Hagiwara:1980nv} where the analogous case of quarkonium decay was studied.
(The quarkonium annihilation beyond leading order was calculated 
earlier in ref.~\cite{Barbieri:1979be}, which
regulated infrared divergences and mass 
singularities using a gluon mass instead of dimensional regularization.)
The $S$-wave stoponium decay width is related to the low-velocity
($v\rightarrow 0$) limit of the stop-anti-stop annihilation cross-section by:
\beq
\Gamma (\eta_{\tilde{t}} \rightarrow X) = 
v\sigma (\tilde{t}_1 \tilde{t}^*_1 \rightarrow X) |\Psi(0)|^2,
\label{eq:Gammasigma}
\eeq
where $\Psi(0)$ is the bound-state wavefunction at the origin.
[This is often expressed instead in terms of the radial wavefunction
at the origin, $R(0) = \sqrt{4\pi}\Psi(0)$.]
Here,
$v$ is the relative velocity of the squarks 
in the center-of-momentum frame.
(The same formula (\ref{eq:Gammasigma})
holds for excited states with 0 angular momentum. 
Obtaining the decay widths of higher angular momentum stoponium
states would require keeping contributions at higher order in 
$v$.)
For diphoton and hadronic final states, the cross-section on the 
right-hand side of eq.~(\ref{eq:Gammasigma}) is in turn related by the 
optical theorem to the imaginary part of the amplitude for 
$\tilde t_1 
\tilde t_1^* \rightarrow \tilde t_1 \tilde t_1^*$ through two-particle 
and 
three-particle cuts. For the $h^0h^0$ final state, 
we find it easier to just 
calculate the radiative corrections to the decay directly. In both cases, 
we work in Feynman gauge, and regulate amplitudes using dimensional 
regularization in $d=4-2\epsilon$ dimensions. Ultraviolet divergences are 
indicated separately by writing $1/\epsUV$, while infrared divergences 
and mass singularities are indicated by $1/\epsIR$ and $1/\epsIR^2$ for 
the pole terms. The top-squark propagator is renormalized on-shell, and 
the QCD gauge coupling will be renormalized in the \MSbar scheme.

\begin{figure}[!tb]
\begin{minipage}[]{0.65\linewidth}
\caption{\label{fig:vdiv} Diagrams contributing to 
$\tilde t_1 \tilde t^*_1 \rightarrow X$
with gluon exchange between the initial state squarks, leading to Coulomb
$1/v$ singularities of the type in eq.~(\ref{eq:vdiv}).}
\end{minipage}
\begin{minipage}[]{0.34\linewidth}
\begin{center}
\begin{picture}(70,70)(-35,-35)
\DashLine(-36,30)(0,10){4}
\DashLine(-36,-30)(0,-10){4}
\Line(20,10)(54,28)
\Line(20,-10)(54,-28)
\Photon(-18,20)(-18,-20){-2.1}{5.5}
\GCirc(10,0){14.14}{0.85}
\Text(-42,22)[]{$\tilde{t}_1$}
\Text(-42,-22)[]{$\tilde{t}_1^*$}
\rText(36,0)[c][r]{$\ldots$}
\end{picture}
\end{center}
\end{minipage}
\end{figure}
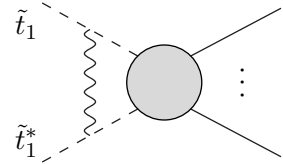

An important issue that arises in all calculations of this type
is that $\sigma (\tilde t_1 \tilde t_1^* \rightarrow X)$
obtains contributions that are divergent as $v\rightarrow 0$ due to the
exchange of massless gluons in diagrams of the form shown in 
figure \ref{fig:vdiv}.
The relevant next-to-leading order contribution in QCD is related 
to the leading order contribution by
\beq
\Delta \sigma^{(1)} (\tilde{t}_1 \tilde{t}_1^{*} \rightarrow X)
\,=\, 
\Bigl [\frac{\pi \alpha_S}{v} C_F + {\cal O}(v^0) \Bigr ]\,
 \sigma^{(0)} (\tilde{t}_1 \tilde{t}_1^{*} \rightarrow X) ,
\label{eq:vdiv}
\eeq
where $C_F$ is the quadratic Casimir invariant, $4/3$ for $SU(3)$.
This Coulomb singularity can be absorbed into the definition of the
bound state wave-function $\Psi(0)$.
Alternatively, since it is universal in character, it cancels when one 
considers
branching ratio observables. This provides a useful test of the calculation.
%%%%%%%%%%%%%%%%%%%%%%%%%%%
%%%%%%%%%%%%%%%%%%%%%%%%%%%%%%%%%%%%%%%%%%%%%%

The rest of this paper is organized as follows. In section 
\ref{sec:hadrons}, we find the next-to-leading order QCD corrections to 
stoponium 
annihilations into $gg$ and $\gamma\gamma$ final states.
Section \ref{sec:hh} discusses the one-loop QCD corrections for
the $h^0h^0$ final state. In section \ref{sec:numerical},
we discuss the numerical impact of these results. Section
\ref{sec:outlook} contains some concluding remarks.

%%%%%%%%%%%%%%%%%%%%%%%%%%%%%%%%%%%%%%%%%%%%%%%%%%%%%%%%%%%%%%%%%%%%%%%%%%%%%%
\section{Decays to hadrons and to photons\label{sec:hadrons}}
\setcounter{footnote}{1}
\setcounter{equation}{0}

In this section, we calculate the next-to-leading order QCD corrections 
to the $gg$ and $\gamma\gamma$ partial widths.  
We use the cut method derived from the optical theorem, 
which allows the 
direct computation of the squared 
amplitude and avoids having to square a matrix element involving many 
terms.  To calculate the amplitude for an arbitrary process $X 
\rightarrow Y$, draw all of the diagrams involving $X \rightarrow X$ 
scattering to the desired order, then cut the diagrams through the 
propagators that correspond to the desired final state $Y$ (for 
$\tilde{t} \tilde{t}^* \rightarrow gg$ at leading order, we have the 
three diagrams in figure \ref{fig:tree}).  These cut propagators are put 
on mass-shell, using the appropriate Feynman rule corresponding to 
external particles (see Figure \ref{fig:feynmanrules}).  Then the sum of 
all cut diagrams, multiplied by an extra factor of $-1$ and 
summed and averaged over spins and colors as appropriate, is denoted by 
$\mathcal{M}_{\rm cut}$ and equals the 
squared amplitude of $X \rightarrow Y$.
%%%%%%%%%%%%%%%%%%%%%%%%%%%%%%%%%%%%%%%%%%%%%%%%%%%%%%%%%%%%%%
\begin{figure}[!tb]
\begin{center}
\begin{picture}(80,60)(0,0)
\Text(35,33)[]{{{$k$}}}
\Text(4,33)[]{{{$j$}}}
\Text(15,48)[]{{{$p$}}}
\DashArrowLine(0,40)(40,40){3}
\Text(48,40)[l]{={\Large{$\frac{i \delta^j_k}{p^2 - m^2}$}}}
\end{picture}
~~~~~~~~
\begin{picture}(80,60)(0,0)
\Text(35,33)[]{{{$k$}}}
\Text(4,33)[]{{{$j$}}}
\Text(10,48)[]{{{$p$}}}
\GBox(19.5,50)(20.5,30){0}
\DashArrowLine(0,40)(40,40){3}
\Text(48,40)[l]{={{$\delta^j_k$}}}
\end{picture}
~~~~~
\begin{picture}(80,60)(0,0)
\Text(35,33)[]{$k$}
\Text(4,33)[]{$j$}
\Text(15,48)[]{$p$}
\ArrowLine(0,40)(40,40)
\Text(48,40)[l]{={\Large{$\frac{i\left(p\!\!\!/\,  + m \right) \delta^j_k}{p^2 -m^2}$}}}
\end{picture}
~~~~~~~~~~~~~~
\begin{picture}(95,60)(0,0)
\Text(35,33)[]{{{$k$}}}
\Text(4,33)[]{{{$j$}}}
\Text(10,48)[]{{{$p$}}}
\GBox(19.5,50)(20.5,30){0}
\ArrowLine(0,40)(40,40)
\Text(48,40)[l]{={{$(p\!\!\!/ +\! m) \delta^j_k$}}}
\end{picture}

\begin{picture}(80,60)(0,0)
\Text(35,33)[]{{{$\nu ,b$}}}
\Text(4,33)[]{{{$\mu ,a$}}}
\Text(15,48)[]{{{$p$}}}
\Photon(0,40)(40,40){1.5}{5.5}
\Text(48,40)[l]{={{$-i\delta^{ab}\frac{g_{\mu \nu} }{p^2}$}}}
\end{picture}
~~~~~~~~~~~~~~
\begin{picture}(87,60)(0,0)
\Text(35,33)[]{{{$\nu ,b$}}}
\Text(4,33)[]{{{$\mu ,a$}}}
\Text(10,48)[]{{{$p$}}}
\GBox(19.5,50)(20.5,30){0}
\Photon(0,40)(40,40){1.5}{5.5}
\Text(48,40)[l]{={{$-g_{\mu \nu} \delta^{ab}$}}}
\end{picture}
~~~~~~~~~~
\begin{picture}(80,60)(0,0)
\Text(35,33)[]{{{$b$}}}
\Text(4,33)[]{{{$a$}}}
\Text(10,48)[]{{{$p$}}}
\DashArrowLine(0,40)(40,40){1}
\Text(48,40)[l]{={\Large{$\frac{i \delta^{ab}}{p^2}$}}}
\end{picture}
~~~~~~~
\begin{picture}(80,60)(0,0)
\Text(35,33)[]{{{$b$}}}
\Text(4,33)[]{{{$a$}}}
\Text(10,48)[]{{{$p$}}}
\GBox(19.5,50)(20.5,30){0}
\DashArrowLine(0,40)(40,40){1}
\Text(48,40)[l]{={{$\delta^{ab}$}}}
\end{picture}

\begin{picture}(105,60)(0,0)
\DashArrowLine(0,60)(21,40){3}
\DashArrowLine(21,40)(0,20){3}
\Text(34,48)[]{{{$\mu ,a$}}}
\Photon(21,40)(42,40){2.0}{4}
\Text(9,72)[t]{{{$p, j$}}}
\Text(9,11)[b]{{{$q, k$}}}
\Text(52,40)[l]{$= ig_3 (p + q)^{\mu}T_k^{a j}$}
\end{picture}
~~~~~~~~~~~~~~~~~
\begin{picture}(105,60)(0,0)
\ArrowLine(0,60)(21,40)
\ArrowLine(21,40)(0,20)
\Text(34,48)[]{{{$\mu ,a$}}}
\Photon(21,40)(44,40){2.0}{4}
\Text(9,72)[t]{{{$j$}}}
\Text(9,11)[b]{{{$k$}}}
\Text(52,40)[l]{$=ig_3 \gamma^{\mu}T_k^{a j}$}
\end{picture}
~~~~~~~~~~~~~
\begin{picture}(105,60)(0,0)
\DashArrowLine(0,60)(21,40){1}
\DashArrowLine(21,40)(0,20){1}
\Text(34,49.5)[]{{{$\mu ,b$}}}
\Photon(21,40)(44,40){2.0}{4}
\Text(9,69)[t]{{{$p, c$}}}
\Text(9,11)[b]{{{$q, a$}}}
\Text(52,40)[l]{$=-g_3 f^{abc}q^{\mu}$}
\end{picture}

\begin{picture}(120,60)(0,0)
\DashArrowLine(0,60)(30,40){3}
\DashArrowLine(30,40)(0,20){3}
\Text(47,68)[t]{{{$\mu ,a$}}}
\Text(47,12)[b]{{{$\nu ,b$}}}
\Photon(30,40)(60,60){1.5}{5}
\Photon(30,40)(60,20){1.5}{5}
\Text(10,72)[t]{{{$p, j$}}}
\Text(10,12)[b]{{{$q, k$}}}
\Text(65,40)[l]{$=ig_3^2 \{ T^a,T^b \}_k^j g^{\mu\nu}$}
\end{picture}
~~~~~~~~~~~~~~~~~~~~~
\begin{picture}(170,60)(0,0)
\DashArrowLine(0,60)(30,40){3}
\DashArrowLine(0,20)(30,40){3}
\Text(45,63)[t]{{{$k$}}}
\Text(45,15)[b]{{{$l$}}}
\DashArrowLine(30,40)(60,60){3}
\DashArrowLine(30,40)(60,20){3}
\Text(15,63)[t]{{{$i$}}}
\Text(15,15)[b]{{{$j$}}}
\Text(65,40)[l]{$=-i g_3^2(T^{ai}_k T^{aj}_l + T^{aj}_k T^{ai}_l)$}
\end{picture}

\begin{picture}(260,64)(0,0)
\Photon(0,60)(30,40){1.5}{4.5}
\Photon(0,20)(30,40){1.5}{4.5}
\Photon(30,40)(60,40){1.5}{4.5}
\Text(15,60)[]{{{$\mu ,a$}}}
\Text(0,50)[]{{{$k$}}}
\Text(0,30)[]{{{$q$}}}
\Text(50,33.5)[]{{{$p$}}}
\Text(15,20)[]{{{$\rho ,c$}}}
\Text(45,50)[]{{{$\nu ,b$}}}
\Text(68,40)[l]{$=g_3 
f^{abc}[g^{\mu\nu}(k - p)^{\rho} + g^{\nu\rho}(p - q)^{\mu} 
+ g^{\rho\mu}(q -k)^{\nu}]$} 
\LongArrow(5,48)(12.5,43)
\LongArrow(5,32)(12.5,37)
\LongArrow(45,35)(37.5,35)
\end{picture}

\begin{picture}(200,36)(0,28)
\Photon(0,60)(30,40){-1.5}{4.5}
\Photon(0,20)(30,40){-1.5}{4.5}
\Photon(30,40)(60,60){1.5}{4.5}
\Photon(30,40)(60,20){1.5}{4.5}
%\Vertex(30,41){2}
\Text(15,63)[]{{{$\mu ,a$}}}
\Text(15,19)[]{{{$\rho ,c$}}}
\Text(45,62)[]{{{$\nu ,b$}}}
\Text(44,18)[]{{{$\sigma ,d$}}}
\Text(68,60)[l]{$= -ig_3^2
[f^{abe}f^{cde}(g^{\mu\rho}g^{\nu\sigma} - g^{\mu\sigma}g^{\nu\rho})$}
\Text(86,40)[l]{$+ f^{ace}f^{bde}(g^{\mu\nu}g^{\rho\sigma} - g^{\mu\sigma}g^{\nu\rho})$}
\Text(86,20)[l]{$+ f^{ade}f^{bce}(g^{\mu\nu}g^{\rho\sigma} - g^{\mu\rho}g^{\nu\sigma})]$}
\end{picture}
\end{center} 
\caption{The Feynman rules used in the calculation of section 
\ref{sec:hadrons}.  Dashed lines refer to scalar particles, solid lines 
are for fermions, wavy lines are for massless gauge (vector) bosons, and 
dotted lines are for the corresponding ghosts.  Dark bars indicate 
propagator cuts.  
QED vertices are obtained by 
the replacement $g_3 T_k^{aj} \rightarrow Q e \delta_k^j$ and
$f^{abc} \rightarrow 0$.  For the 
four-point scalar vertex, our effective theory neglects Yukawa and 
electroweak couplings that occur in the full MSSM. 
\label{fig:feynmanrules}}
\end{figure}
%%%%%%%%%%%%%%%%%%%%%%%%%%%%%%%%%%%%%%%%%%%%%%%%

In order to get the partial widths, the contributions to 
${\mathcal M}_{\rm cut}$ must be integrated 
over $d$-dimensional Lorentz-invariant phase space. The individual 
contribution of a single cut diagram to the cross-section is
\beq
\Delta \sigma & = & \frac{1}{4 E_A E_B v} 
\int 
\mathcal{M}_{\rm cut} 
\biggl (\prod_f{ 
\mu^{2\epsilon} \frac{d^d k_f}{(2 \pi)^{d-1}} \delta(k_f^2 - m_f^2) 
\theta(k_f^0)} 
\biggr ) 
(2\pi)^d 
\delta^{(d)}(p_A + p_B - \sum_f k_f) 
\nonumber \\
& \equiv & \frac{1}{4 E_A E_B v} \int{\mathcal{M}_{\rm cut} 
\ {\rm dLIPS}_N}.
\label{eq:deltasigma}
\eeq
In equation (\ref{eq:deltasigma}), 
the labels $A$ and $B$ are for the initial-state and $f$ 
for the final-state momentum four-vectors, $v$ is the relative velocity 
of the 
initial-state particles, and $\int{{\rm dLIPS}_N}$ is the integral over 
$N$-body Lorentz-invariant phase space.  Since we are calculating the 
annihilation of a bound state, we multiply both sides by the relative 
velocity and set $E_A = E_B = m_{\tilde t_1}$ as the relative velocity 
goes to zero. Therefore,
\beq
v \Delta \sigma = \frac{1}{4 m_{\tilde t_1}^2} 
\int{\mathcal{M}_{\rm cut} \ {\rm dLIPS}_N}.
\label{eq:deltavsigma}
\eeq
Adding up all of the terms from the appropriate cut diagrams gives the 
total cross-section multiplied by the 
relative velocity, $v \sigma$, which is 
related to the partial width in equation (\ref{eq:Gammasigma}).

At tree-level, the cross section for the annihilation of $\tilde{t}_1 
\tilde{t}_1^{*}$ into a gluon-gluon final state in $d = 4 - 2 \epsilon$ 
dimensions is
\beq
v \sigma^{(0)} (\tilde{t}_1 \tilde{t}_1^{*} \rightarrow gg) = 
\frac{\pi \widehat \alpha_S^2}{2m_{\tilde{t}_1}^2} 
\biggl( \frac{N_c^2 - 1}{N_c} \biggr) 
\frac{\Gamma (2-\epsilon)}{\Gamma (2 - 2\epsilon)}  
\biggl ( \frac{\pi \mu^4}{m_{\tilde{t}_1}^2} \biggr )^{\epsilon},
\eeq
with $\widehat \alpha_S = (\hat g_3^2/4 \pi)\mu^{-2 \epsilon}$, where
$\hat g_3$ is the bare QCD coupling and $\mu$ is the regularization
mass. The 
diagrams $a$, 
$b$, and $c$ in figure {\ref{fig:tree}} contribute to this result in the 
ratio $0:-1:(2-\epsilon)$.
%%%%%%%%%%%%%%%%%%%%%%%%%%%%%%%%%%%%%%%%%%%%%%%%%%%%%%%%%%%%%%%%%%%%%%%%%%%%
%
\begin{figure}[!tb]
\begin{picture}(63,70)(0,0)
\DashLine(0,60)(21,60){3}
\DashLine(21,60)(21,20){3}
\DashLine(0,20)(21,20){3}
\DashLine(42,60)(63,60){3}
\DashLine(42,20)(42,60){3}
\DashLine(42,20)(63,20){3}
\Photon(21,60)(42,60){1.5}{4.5}
\Photon(21,20)(42,20){1.5}{4.5}
\GBox(31,15)(32,65){0}
\Text(31.5,0)[b]{a}
\end{picture}
\hspace{0.25in}
\begin{picture}(63,70)(0,0)
\DashLine(0,60)(21,60){3}
\DashLine(21,60)(21,20){3}
\DashLine(0,20)(21,20){3}
\DashLine(42,40)(63,60){3}
\DashLine(42,40)(63,20){3}
\Photon(21,60)(42,40){1.5}{5.5}
\Photon(21,20)(42,40){-1.5}{5.5}
\GBox(31,15)(32,65){0}
\Text(31.5,0)[b]{b}
\end{picture}
\hspace{0.25in}
\begin{picture}(63,70)(0,0)
\DashLine(42,40)(63,60){3}
\DashLine(42,40)(63,20){3}
\DashLine(0,20)(21,40){3}
\DashLine(0,60)(21,40){3}
\PhotonArc(31.5,40)(10.5,0,360){1.3}{11.5}
\GBox(31,15)(32,65){0}
\Text(31.5,0)[b]{c}
\end{picture}
\caption{\label{fig:tree}
The diagrams whose imaginary parts contribute to the
annihilation of $\tilde{t}_1\tilde{t}_1^{*}$ 
into gauge bosons at leading order.  
The dark bars indicate where the 
diagrams have been cut. Diagrams related by permutations 
of external lines and arrow-reversal are not distinguished.}  
\end{figure}
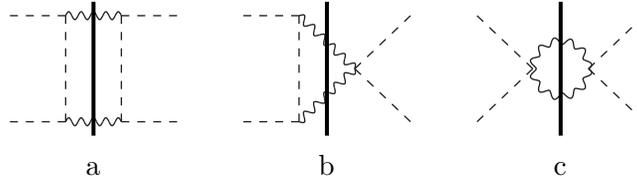
%%%%%%%%%%%%%%%%%%%%%%%%%%%%%%%%%%%%%%%%%%%%%%%%%%%%%%%%%%%%

The tree-level diphoton cross section can be 
obtained by the replacement $\hat g_3 T_k^{aj} \rightarrow Q e 
\delta_k^j$ in the $gg$ result.  For this final state, 
\beq
v \sigma^{(0)} (\tilde{t}_1 \tilde{t}_1^{*} \rightarrow \gamma \gamma) = 
\frac{2 N_c \pi Q_s^4 \alpha^2}{m_{\tilde{t}_1}^2} 
\frac{\Gamma (2-\epsilon)}{\Gamma (2 - 2\epsilon)}  
\biggl( \frac{\pi \mu^4}{m_{\tilde{t}_1}^2} \biggr )^{\epsilon},
\eeq
where $Q_s = +2/3$ is the charge of the squark, and
$\alpha = e^2/4 \pi$.
Therefore, the leading-order decay rates are
\beq
\Gamma^{(0)} \left( \eta_{\tilde{t}} \rightarrow gg \right) & = & 
\frac{16\pi}{3}  \alpha_S^2 
\frac{|\Psi(0)|^2}{m_{\eta_{\tilde{t}}}^2} 
,
\\
 & & \nonumber \\ 
\Gamma^{(0)} \left( \eta_{\tilde{t}} \rightarrow \gamma \gamma \right) & = &
\frac{128\pi}{27}  \alpha^2 
\frac{|\Psi(0)|^2}{m_{\eta_{\tilde{t}}}^2}
,
\label{eq:Gamgamgam}
\eeq
where we have replaced the bare coupling $\widehat \alpha_S$ with the
renormalized coupling $\alpha_S$, since they are equal at leading order.
Taking the ratio of these partial widths eliminates the bound 
state wavefunction and produces the simple leading-order result
\beq
R^{(0)} \equiv 
\frac{\Gamma^{(0)} \left( \eta_{\tilde{t}} \rightarrow 
\gamma\gamma\right)}{\Gamma^{(0)} 
\left( \eta_{\tilde{t}} \rightarrow gg \right)} = 
\frac{8 \alpha^2}{9 \alpha_S^2}.
\label{eq:Rzero}
\eeq

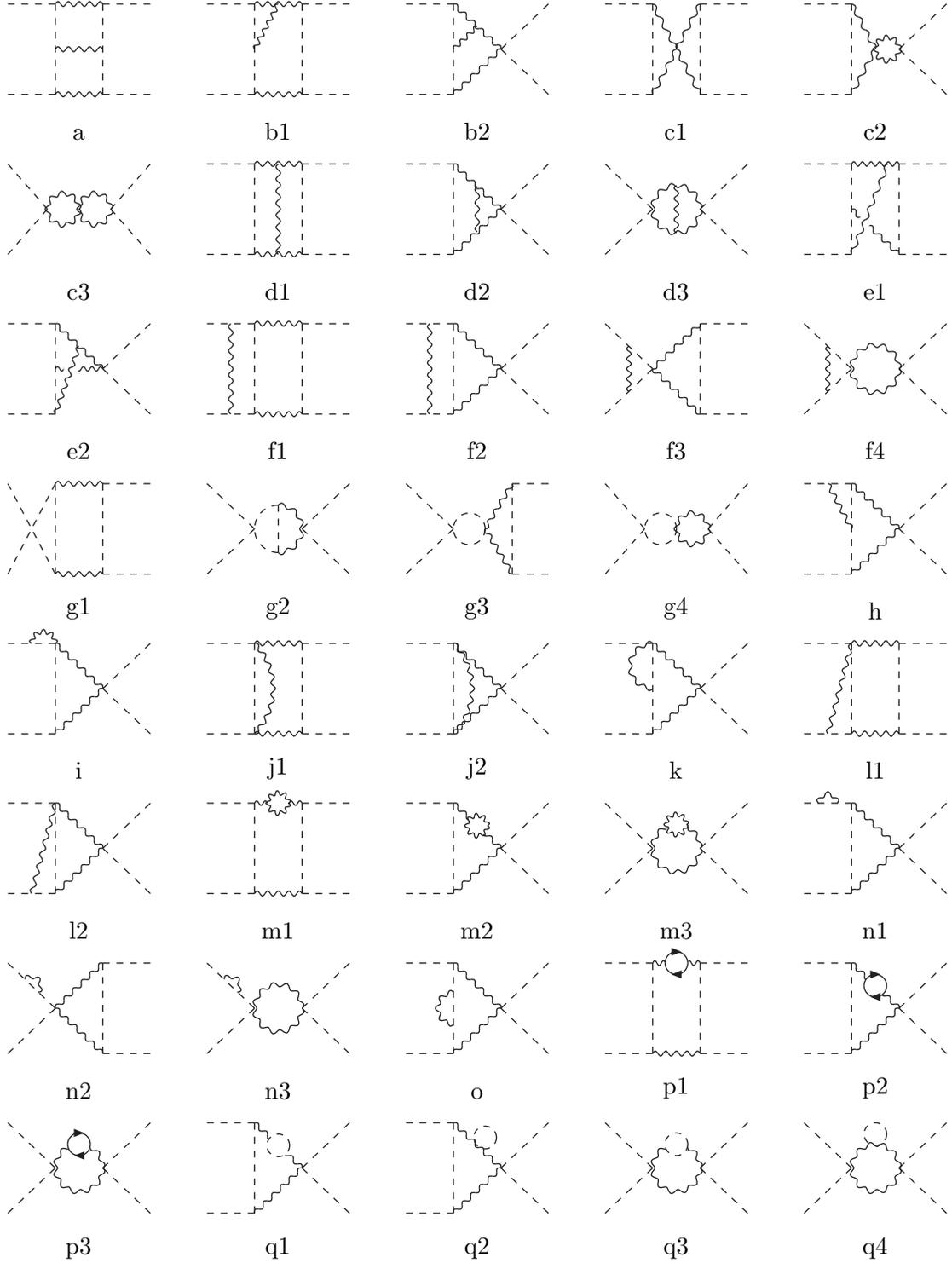
\begin{figure}[p]
\begin{picture}(63,70)(0,0)
\DashLine(0,60)(21,60){3}
\DashLine(21,60)(21,20){3}
\DashLine(0,20)(21,20){3}
\DashLine(42,60)(63,60){3}
\DashLine(42,20)(42,60){3}
\DashLine(42,20)(63,20){3}
\Photon(21,60)(42,60){1.1}{4.5}
\Photon(21,40)(42,40){1.1}{4.5}
\Photon(21,20)(42,20){1.1}{4.5}
\Text(31.5,0)[b]{a}
\end{picture}
\hspace{0.25in}
\begin{picture}(63,70)(0,0)
\DashLine(0,60)(21,60){3}
\DashLine(21,60)(21,20){3}
\DashLine(0,20)(21,20){3}
\DashLine(42,60)(63,60){3}
\DashLine(42,20)(42,60){3}
\DashLine(42,20)(63,20){3}
\Photon(21,60)(42,60){1.1}{4.5}
\Photon(21,40)(31.5,60){1}{5.5}
\Photon(21,20)(42,20){1.1}{4.5}
\Text(31.5,0)[b]{b1}
\end{picture}
\hspace{0.25in}
\begin{picture}(63,70)(0,0)
\DashLine(0,60)(21,60){3}
\DashLine(21,60)(21,20){3}
\DashLine(0,20)(21,20){3}
\DashLine(42,40)(63,60){3}
\DashLine(42,40)(63,20){3}
\Photon(21,60)(42,40){1}{5.5}
\Photon(21,40)(31.5,50){1}{3.5}
\Photon(21,20)(42,40){1}{5.5}
\Text(31.5,0)[b]{b2}
\end{picture}
\hspace{0.25in}
\begin{picture}(63,70)(0,0)
\DashLine(0,60)(21,60){3}
\DashLine(21,60)(21,20){3}
\DashLine(0,20)(21,20){3}
\DashLine(42,60)(63,60){3}
\DashLine(42,20)(42,60){3}
\DashLine(42,20)(63,20){3}
\Photon(21,20)(31.5,40){1}{3.5}
\Photon(21,60)(31.5,40){1}{3.5}
\Photon(31.5,40)(42,60){1}{3.5}
\Photon(31.5,40)(42,20){1}{3.5}
%\Vertex(31.5,40){1.25}
\Text(31.5,0)[b]{c1}
\end{picture}
\hspace{0.25in}
\begin{picture}(63,70)(0,0)
\DashLine(0,60)(21,60){3}
\DashLine(21,60)(21,20){3}
\DashLine(0,20)(21,20){3}
\DashLine(42,40)(63,60){3}
\DashLine(42,40)(63,20){3}
\Photon(21,20)(31.5,40){1}{3.5}
\Photon(21,60)(31.5,40){1}{3.5}
%\Vertex(31.5,40){1.25}
\PhotonArc(36.75,40)(5.25,0,360){1}{8.5}
\Text(31.5,0)[b]{c2}
\end{picture}

\begin{picture}(63,70)(0,0)
\DashLine(45.5,40)(63,60){3}
\DashLine(45.5,40)(63,20){3}
\DashLine(0,20)(17.5,40){3}
\DashLine(0,60)(17.5,40){3}
\PhotonArc(24.5,40)(7,0,360){1}{8.5}
\PhotonArc(38.5,40)(7,0,360){1}{8.5}
%\Vertex(31.5,40){1.25}
\Text(31.5,0)[b]{c3}
\end{picture}
\hspace{0.25in}
\begin{picture}(63,70)(0,0)
\DashLine(0,60)(21,60){3}
\DashLine(21,60)(21,20){3}
\DashLine(0,20)(21,20){3}
\DashLine(42,60)(63,60){3}
\DashLine(42,20)(42,60){3}
\DashLine(42,20)(63,20){3}
\Photon(21,60)(42,60){1.1}{4.5}
\Photon(31.5,20)(31.5,60){1}{7.5}
\Photon(21,20)(42,20){1.1}{4.5}
\Text(31.5,0)[b]{d1}
\end{picture}
\hspace{0.25in}
\begin{picture}(63,70)(0,0)
\DashLine(0,60)(21,60){3}
\DashLine(21,60)(21,20){3}
\DashLine(0,20)(21,20){3}
\DashLine(42,40)(63,60){3}
\DashLine(42,40)(63,20){3}
\Photon(21,60)(42,40){1}{5.5}
\Photon(31.5,30)(31.5,50){1}{3.5}
\Photon(21,20)(42,40){1}{5.5}
\Text(31.5,0)[b]{d2}
\end{picture}
\hspace{0.25in}
\begin{picture}(63,70)(0,0)
\DashLine(42,40)(63,60){3}
\DashLine(42,40)(63,20){3}
\DashLine(0,20)(21,40){3}
\DashLine(0,60)(21,40){3}
\PhotonArc(31.5,40)(10.5,0,360){1}{11.5}
\Photon(31.5,29.5)(31.5,50.5){1}{4.5}
\Text(31.5,0)[b]{d3}
\end{picture}
\hspace{0.25in}
\begin{picture}(63,70)(0,0)
\DashLine(0,60)(21,60){3}
\DashLine(21,60)(21,20){3}
\DashLine(0,20)(21,20){3}
\DashLine(42,60)(63,60){3}
\DashLine(42,20)(42,60){3}
\DashLine(42,20)(63,20){3}
\Photon(21,60)(42,60){1}{5.5}
\Photon(21,40)(42,20){1}{5.5}
\COval(26.79,34.48)(2.5,2.5)(0){White}{White}
\Photon(21,20)(37,60){1}{6.5}
\Text(31.5,0)[b]{e1}
\end{picture}

\begin{picture}(63,70)(0,0)
\DashLine(0,60)(21,60){3}
\DashLine(21,60)(21,20){3}
\DashLine(0,20)(21,20){3}
\DashLine(42,40)(63,60){3}
\DashLine(42,40)(63,20){3}
\Photon(21,60)(42,40){1}{5.5}
\Photon(21,40)(42,40){1}{5.5}
\COval(28,40)(2.5,3)(0){White}{White}
\Photon(21,20)(31.5,50){1}{6.5}
\Text(31.5,0)[b]{e2}
\end{picture}
\hspace{0.25in}
\begin{picture}(63,70)(0,0)
\DashLine(0,60)(21,60){3}
\DashLine(21,60)(21,20){3}
\DashLine(0,20)(21,20){3}
\DashLine(42,60)(63,60){3}
\DashLine(42,20)(42,60){3}
\DashLine(42,20)(63,20){3}
\Photon(21,60)(42,60){1.1}{4.5}
\Photon(10.5,20)(10.5,60){1}{7.5}
\Photon(21,20)(42,20){1.1}{4.5}
\Text(31.5,0)[b]{f1}
\end{picture}
\hspace{0.25in}
\begin{picture}(63,70)(0,0)
\DashLine(0,60)(21,60){3}
\DashLine(21,60)(21,20){3}
\DashLine(0,20)(21,20){3}
\DashLine(42,40)(63,60){3}
\DashLine(42,40)(63,20){3}
\Photon(21,60)(42,40){1}{5.5}
\Photon(10.5,20)(10.5,60){1}{7.5}
\Photon(21,20)(42,40){1}{5.5}
\Text(31.5,0)[b]{f2}
\end{picture}
\hspace{0.25in}
\begin{picture}(63,70)(0,0)
\DashLine(42,60)(63,60){3}
\DashLine(42,60)(42,20){3}
\DashLine(42,20)(63,20){3}
\DashLine(21,40)(0,60){3}
\DashLine(21,40)(0,20){3}
\Photon(42,60)(21,40){1}{5.5}
\Photon(10.5,30)(10.5,50){1}{4.5}
\Photon(42,20)(21,40){1}{5.5}
\Text(31.5,0)[b]{f3}
\end{picture}
\hspace{0.25in}
\begin{picture}(63,70)(0,0)
\DashLine(42,40)(63,60){3}
\DashLine(42,40)(63,20){3}
\DashLine(0,20)(21,40){3}
\DashLine(0,60)(21,40){3}
\Photon(10.5,30)(10.5,50){1}{4.5}
\PhotonArc(31.5,40)(10.5,0,360){1}{11.5}
\Text(31.5,0)[b]{f4}
\end{picture}

\begin{picture}(63,70)(0,0)
\DashLine(42,60)(63,60){3}
\DashLine(42,60)(42,20){3}
\DashLine(42,20)(63,20){3}
%\Vertex(10.5,40){1.25}
\DashLine(0,20)(21,60){3}
\DashLine(0,60)(21,20){3}
\DashLine(21,60)(21,20){3}
\Photon(21,60)(42,60){1.1}{4.5}
\Photon(21,20)(42,20){1.1}{4.5}
\Text(31.5,0)[b]{g1}
\end{picture}
\hspace{0.25in}
\begin{picture}(63,70)(0,0)
\DashLine(42,40)(63,60){3}
\DashLine(42,40)(63,20){3}
\DashLine(0,20)(21,40){3}
\DashLine(0,60)(21,40){3}
\DashCArc(31.5,40)(10.5,90,270){3}
\DashLine(31.5,29.5)(31.5,50.5){3}
\PhotonArc(31.5,40)(10.5,-90,90){1}{5.5}
%\Vertex(21,40){1.25}
\Text(31.5,0)[b]{g2}
\end{picture}
\hspace{0.25in}
\begin{picture}(63,70)(0,0)
\DashLine(0,60)(21,40){3}
\DashLine(0,20)(21,40){3}
\DashLine(47,60)(63,60){3}
\DashLine(47,60)(47,20){3}
\DashLine(47,20)(63,20){3}
\Photon(47,60)(35,40){1}{4.5}
\Photon(47,20)(35,40){1}{4.5}
\DashCArc(28,40)(7,0,360){3}
%\Vertex(21,40){1.25}
\Text(31.5,0)[b]{g3}
\end{picture}
\hspace{0.25in}
\begin{picture}(63,70)(0,0)
\DashLine(45.5,40)(63,60){3}
\DashLine(45.5,40)(63,20){3}
\DashLine(0,20)(17.5,40){3}
\DashLine(0,60)(17.5,40){3}
\PhotonArc(38.5,40)(7,0,360){1}{8.5}
\DashCArc(24.5,40)(7,0,360){3}
%\Vertex(17.5,40){1.25}
\Text(31.5,0)[b]{g4}
\end{picture}
\hspace{0.25in}
\begin{picture}(63,70)(0,0)
\DashLine(0,60)(21,60){3}
\DashLine(21,60)(21,20){3}
\DashLine(0,20)(21,20){3}
\DashLine(42,40)(63,60){3}
\DashLine(42,40)(63,20){3}
\Photon(21,60)(42,40){1}{5.5}
\Photon(21,20)(42,40){1}{5.5}
\Photon(10.5,60)(21,40){1}{4.5}
\Text(31.5,0)[b]{h}
\end{picture}

\begin{picture}(63,70)(0,0)
\DashLine(0,60)(21,60){3}
\DashLine(21,60)(21,20){3}
\DashLine(0,20)(21,20){3}
\DashLine(42,40)(63,60){3}
\DashLine(42,40)(63,20){3}
\Photon(21,60)(42,40){1}{5.5}
\Photon(21,20)(42,40){1}{5.5}
\PhotonArc(15.75,60)(5.25,0,180){1}{4.5}
%\Vertex(21,60){1.25}
\Text(31.5,0)[b]{i}
\end{picture}
\hspace{0.25in}
\begin{picture}(63,70)(0,0)
\DashLine(0,60)(21,60){3}
\DashLine(21,60)(21,20){3}
\DashLine(0,20)(21,20){3}
\DashLine(42,60)(63,60){3}
\DashLine(42,20)(42,60){3}
\DashLine(42,20)(63,20){3}
\Photon(21,60)(42,60){1.1}{4.5}
\Photon(21,20)(42,20){1.1}{4.5}
\PhotonArc(1,40)(28.2843,-45,45){1}{8.5}
%\Vertex(21,60){1.25}
%\Vertex(21,20){1.25}
\Text(31.5,0)[b]{j1}
\end{picture}
\hspace{0.25in}
\begin{picture}(63,70)(0,0)
\DashLine(0,60)(21,60){3}
\DashLine(21,60)(21,20){3}
\DashLine(0,20)(21,20){3}
\DashLine(42,40)(63,60){3}
\DashLine(42,40)(63,20){3}
\Photon(21,60)(42,40){1}{5.5}
\Photon(21,20)(42,40){1}{5.5}
\PhotonArc(1,40)(28.2843,-45,45){1}{8.5}
%\Vertex(21,60){1.25}
%\Vertex(21,20){1.25}
\Text(31.5,0)[b]{j2}
\end{picture}
\hspace{0.25in}
\begin{picture}(63,70)(0,0)
\DashLine(0,60)(21,60){3}
\DashLine(21,60)(21,20){3}
\DashLine(0,20)(21,20){3}
\DashLine(42,40)(63,60){3}
\DashLine(42,40)(63,20){3}
\Photon(21,60)(42,40){1}{5.5}
\Photon(21,20)(42,40){1}{5.5}
\PhotonArc(21,50)(10,90,270){1}{5.5}
%\Vertex(21,60){1.25}
\Text(31.5,0)[b]{k}
\end{picture}
\hspace{0.25in}
\begin{picture}(63,70)(0,0)
\DashLine(0,60)(21,60){3}
\DashLine(21,60)(21,20){3}
\DashLine(0,20)(21,20){3}
\DashLine(42,60)(63,60){3}
\DashLine(42,20)(42,60){3}
\DashLine(42,20)(63,20){3}
\Photon(21,60)(42,60){1.2}{4.5}
\Photon(21,20)(42,20){1.2}{4.5}
\Photon(10.5,20)(21,60){1}{7.5}
%\Vertex(21,60){1.25}
\Text(31.5,0)[b]{l1}
\end{picture}

\begin{picture}(63,70)(0,0)
\DashLine(0,60)(21,60){3}
\DashLine(21,60)(21,20){3}
\DashLine(0,20)(21,20){3}
\DashLine(42,40)(63,60){3}
\DashLine(42,40)(63,20){3}
\Photon(21,60)(42,40){1}{5.5}
\Photon(21,20)(42,40){1}{5.5}
\Photon(10.5,20)(21,60){1}{7.5}
%\Vertex(21,60){1.25}
\Text(31.5,0)[b]{l2}
\end{picture}
\hspace{0.25in}
\begin{picture}(63,70)(0,0)
\DashLine(0,60)(21,60){3}
\DashLine(21,60)(21,20){3}
\DashLine(0,20)(21,20){3}
\DashLine(42,60)(63,60){3}
\DashLine(42,20)(42,60){3}
\DashLine(42,20)(63,20){3}
\Photon(21,60)(42,60){1}{5.5}
\CCirc(31.5,60){4.5}{White}{White}
\PhotonArc(31.5,60)(4.5,0,360){1}{8.5}
\Photon(21,20)(42,20){1.1}{4.5}
\Text(31.5,0)[b]{m1}
\end{picture}
\hspace{0.25in}
\begin{picture}(63,70)(0,0)
\DashLine(0,60)(21,60){3}
\DashLine(21,60)(21,20){3}
\DashLine(0,20)(21,20){3}
\DashLine(42,40)(63,60){3}
\DashLine(42,40)(63,20){3}
\Photon(21,60)(42,40){1}{5.5}
\Photon(21,20)(42,40){1}{5.5}
\CCirc(31.5,50){4.5}{White}{White}
\PhotonArc(31.5,50)(4.5,0,360){1}{8.5}
\Text(31.5,0)[b]{m2}
\end{picture}
\hspace{0.25in}
\begin{picture}(63,70)(0,0)
\DashLine(42,40)(63,60){3}
\DashLine(42,40)(63,20){3}
\DashLine(0,20)(21,40){3}
\DashLine(0,60)(21,40){3}
\PhotonArc(31.5,40)(10.5,0,360){1}{11.5}
\CCirc(31.5,50.5){4.5}{White}{White}
\PhotonArc(31.5,50.5)(4.5,0,360){1}{8.5}
\Text(31.5,0)[b]{m3}
\end{picture}
\hspace{0.25in}
\begin{picture}(63,70)(0,0)
\DashLine(0,60)(21,60){3}
\DashLine(21,60)(21,20){3}
\DashLine(0,20)(21,20){3}
\DashLine(42,40)(63,60){3}
\DashLine(42,40)(63,20){3}
\Photon(21,60)(42,40){1}{5.5}
\Photon(21,20)(42,40){1}{5.5}
\PhotonArc(10.5,60)(4,0,180){1}{2.5}
\Text(31.5,0)[b]{n1}
\end{picture}

\begin{picture}(63,70)(0,0)
\DashLine(63,60)(42,60){3}
\DashLine(42,60)(42,20){3}
\DashLine(42,20)(63,20){3}
\DashLine(21,40)(0,60){3}
\DashLine(21,40)(0,20){3}
\Photon(21,40)(42,60){1}{5.5}
\Photon(21,40)(42,20){1}{5.5}
\PhotonArc(10.5,50)(4,-45,135){1}{2.5}
\Text(31.5,0)[b]{n2}
\end{picture}
\hspace{0.25in}
\begin{picture}(63,70)(0,0)
\DashLine(42,40)(63,60){3}
\DashLine(42,40)(63,20){3}
\DashLine(0,20)(21,40){3}
\DashLine(0,60)(21,40){3}
\PhotonArc(10.5,50)(4,-45,135){1}{2.5}
\PhotonArc(31.5,40)(10.5,0,360){1}{11.5}
\Text(31.5,0)[b]{n3}
\end{picture}
\hspace{0.25in}
\begin{picture}(63,70)(0,0)
\DashLine(0,60)(21,60){3}
\DashLine(21,60)(21,20){3}
\DashLine(0,20)(21,20){3}
\DashLine(42,40)(63,60){3}
\DashLine(42,40)(63,20){3}
\Photon(21,60)(42,40){1}{5.5}
\Photon(21,20)(42,40){1}{5.5}
\PhotonArc(21,40)(7,90,270){1}{4.5}
\Text(31.5,0)[b]{o}
\end{picture}
\hspace{0.25in}
\begin{picture}(63,70)(0,0)
\DashLine(0,60)(21,60){3}
\DashLine(21,60)(21,20){3}
\DashLine(0,20)(21,20){3}
\DashLine(42,60)(63,60){3}
\DashLine(42,20)(42,60){3}
\DashLine(42,20)(63,20){3}
\Photon(21,60)(42,60){1.1}{5.5}
\Photon(21,20)(42,20){1.1}{4.5}
\CCirc(31.5,60){5}{White}{White}
\ArrowArcn(31.5,60)(5,0,180)
\ArrowArcn(31.5,60)(5,180,360)
\Text(31.5,0)[b]{p1}
\end{picture}
\hspace{0.25in}
\begin{picture}(63,70)(0,0)
\DashLine(0,60)(21,60){3}
\DashLine(21,60)(21,20){3}
\DashLine(0,20)(21,20){3}
\DashLine(42,40)(63,60){3}
\DashLine(42,40)(63,20){3}
\Photon(21,60)(42,40){1}{5.5}
\Photon(21,20)(42,40){1}{5.5}
\CCirc(31.5,50){5}{White}{White}
\ArrowArcn(31.5,50)(5,0,180)
\ArrowArcn(31.5,50)(5,180,360)
\Text(31.5,0)[b]{p2}
\end{picture}

\begin{picture}(63,70)(0,0)
\DashLine(42,40)(63,60){3}
\DashLine(42,40)(63,20){3}
\DashLine(0,20)(21,40){3}
\DashLine(0,60)(21,40){3}
\PhotonArc(31.5,40)(10.5,0,360){1}{11.5}
\CCirc(31.5,50.5){5}{White}{White}
\ArrowArcn(31.5,50.5)(5,0,180)
\ArrowArcn(31.5,50.5)(5,180,360)
\Text(31.5,0)[b]{p3}
\end{picture}
\hspace{0.25in}
\begin{picture}(63,70)(0,0)
\DashLine(0,60)(21,60){3}
\DashLine(21,60)(21,20){3}
\DashLine(0,20)(21,20){3}
\DashLine(42,40)(63,60){3}
\DashLine(42,40)(63,20){3}
\Photon(21,60)(42,40){1}{5.5}
\Photon(21,20)(42,40){1}{5.5}
\CCirc(31.5,50){5}{White}{White}
\DashCArc(31.5,50)(5,0,360){3}
\Text(31.5,0)[b]{q1}
\end{picture}
\hspace{0.25in}
\begin{picture}(63,70)(0,0)
\DashLine(0,60)(21,60){3}
\DashLine(21,60)(21,20){3}
\DashLine(0,20)(21,20){3}
\DashLine(42,40)(63,60){3}
\DashLine(42,40)(63,20){3}
\Photon(21,60)(42,40){1}{5.5}
\Photon(21,20)(42,40){1}{5.5}
\DashCArc(35,53.5)(5,0,360){3}
%\Vertex(31.5,50){1.25}
\Text(31.5,0)[b]{q2}
\end{picture}
\hspace{0.25in}
\begin{picture}(63,70)(0,0)
\DashLine(42,40)(63,60){3}
\DashLine(42,40)(63,20){3}
\DashLine(0,20)(21,40){3}
\DashLine(0,60)(21,40){3}
\PhotonArc(31.5,40)(10.5,0,360){1}{11.5}
\CCirc(31.5,50.5){5}{White}{White}
\DashCArc(31.5,50.5)(5,0,360){3}
\Text(31.5,0)[b]{q3}
\end{picture}
\hspace{0.25in}
\begin{picture}(63,70)(0,0)
\DashLine(42,40)(63,60){3}
\DashLine(42,40)(63,20){3}
\DashLine(0,20)(21,40){3}
\DashLine(0,60)(21,40){3}
\PhotonArc(31.5,40)(10.5,0,360){1}{11.5}
\DashCArc(31.5,55)(5,0,360){3}
%\Vertex(31.5,50.5){1.25}
\Text(31.5,0)[b]{q4}
\end{picture}

\vspace{0.2in}

\caption{\label{fig:diagrams}
The diagrams whose imaginary parts contribute to 
the cross-section for $\tilde t_1 \tilde t_1^* \rightarrow 
gg$ at next-to-leading order, including $ggg$ and $gq\overline q$.  Each 
diagram must be cut in all the 
ways it is 
possible to put the cut propagators simultaneously on-shell, and the 
propagators that are cut indicate the corresponding two- or 
three-particle final state.  Diagrams related by permutations 
of external states and 
arrow-reversal are not shown, nor are the diagrams with ghost loops 
that are needed for each diagram with a gauge boson loop.  We have also 
not shown several diagrams that vanish for all possible cuts.}
\end{figure}

The non-vanishing cut diagrams that correspond to the annihilation of 
$\tilde{t}_1 \tilde{t}_1^*$ into $gg$, $ggg$, and $gq\overline q$ final 
states at next-to-leading order are given in figure 
\ref{fig:diagrams}\footnote{Note that several diagrams not shown in the 
figure vanish because the color indices of three-gluon final states must 
be antisymmetric by charge conjugation invariance. This is because 
$\eta_{\tilde t}$ has $C = +1$, while a final state with $n$ gluons has 
$C = (-1)^{n+n_c}$, where $n_c$ is $1$ ($0$) for antisymmetric 
(symmetric) adjoint color indices \cite{Novikov}.}.%
%%%%%%%%%%%%%%%%%%%%%%%%%%%%%%%%%%%%%%%%%%%%%%%%
%
%%%%%%%%%%%%%%%%%%%%%%%%%%%%%%%%%%%%%%%%%%%%%%%%%%%%%%%%%%%%%%%%%%%%%%%%%
Many of these 
diagrams can be cut in more than one way. In diagrams with three 
cut propagators, there is either real gluon emission or the 
pair-production of quarks\footnote{We include some $gq\overline q$
final state contributions,
even 
though these may be regarded as corrections to 
$q\overline q$ 
final states, which we do not treat here.  The light $q\overline q$ 
partial widths are suppressed by small Yukawa couplings at 
leading order, and 
the $t \overline t$ final state is often strongly suppressed
by kinematics and couplings. However, the
$gq\overline q$ contributions from 3-particle cuts in
diagrams p1, p2, and p3 in figure {\ref{fig:diagrams}}
cancel
large logarithms in the limit 
of small $m_q$
due to the gluon vacuum polarization (2-particle cut)
contributions from the same diagrams.}, and diagrams 
with two cut propagators have one-loop integrals.

In diagrams with three-particle cuts, the principal difficulty is 
integrating the momentum fractions of the final-state particles over 
three-body phase space.  To do this, the phase space integrals can be 
reduced to integrals of the form given in the Appendix of 
ref.~\cite{Hagiwara:1980nv}.  Care must be taken in evaluating diagrams with 
multiple distinct three-propagator cuts.  Diagram d2, for example, has 
two cuts that are not equal.

Evaluation of the two-particle cuts involves expanding the loop integral 
from the virtual gluon in partial fractions to obtain a set of scalar 
integrals, which are well-known.  A complete set of scalar integrals that 
occur in the calculation can be found in ref. \cite{Beenakker:1988bq} 
(for a complete set of divergent and many finite scalar loop integrals, 
see ref. \cite{Ellis:2007qk}).  In contrast with the three-particle cuts, 
the phase space integration is quite easy \cite{Hagiwara:1980nv}.  
Since the cut diagrams do not depend 
on the final-state momentum directions, 
they are proportional to their contributions to the cross-section 
\beq
v \Delta \sigma^{(1)} & = & 
\frac{1}{4 m_{\tilde t_1}^2} \mathcal{M}_{\rm cut} \Phi(2),
\eeq
where
\beq
\Phi(2) 
\equiv 
\int{{\rm dLIPS}_2} 
= \frac{1}{8\pi} \left( \frac{\pi}{m^2_{\tilde t_1}} \right)^{\epsilon} 
\frac{\Gamma(1-\epsilon)}{\Gamma(2-2\epsilon)} .
\label{eq:twobodyvsigma}
\eeq

There is an important simplification that can be made in the massless two-particle 
cuts of diagrams with potential Coulomb singularities
(diagrams f1, f2, f3, and f4 in figure \ref{fig:diagrams}). 
Using the identities \cite{Hagiwara:1980nv}
\beq
\int{\mbox{dLIPS}(P=k_1 + k_2)k_1^{\mu}} & = & \frac{1}{2} P^{\mu} \Phi(2) \\
\int{\mbox{dLIPS}(P=k_1 + k_2)k_1^{\mu}k_1^{\nu}} & = & \frac{1}{4} \left( \frac{d}{d-1} P^{\mu} P^{\nu} - \frac{1}{d-1} 
P^2 g^{\mu \nu} \right) \Phi(2),
\eeq
where $k_1$ and $k_2$ are the final-state 
gluon momentum 4-vectors, $d \equiv 4 - 2 
\epsilon$ 
is the number of spacetime dimensions, and $\Phi(2)$ the integrated 
two-body phase space, it is easy to show that, for integrals performed in 
this calculation, dot products of the particle momentum 4-vectors in the 
numerators of loop integrals cannot contain terms linear in the relative 
velocity $v$. The effect of the $v \rightarrow 0$ divergence comes only 
from the scalar loop integrals in the calculation, and one may set $v=0$ 
everywhere else.

In Tables {\ref{tab:nonpropagator}} and {\ref{tab:propagator}}, the 
contribution $C_{\rm diagram}$ from each diagram to $v 
\sigma^{(1)}(\tilde{t}_1 \tilde{t}_1^{*} \rightarrow gg)$ is given in the 
form
\beq
v \Delta \sigma^{(1)}(\tilde{t}_1 \tilde{t}_1^{*} \rightarrow gg) = 
v\sigma^{(0)}(\tilde{t}_1 \tilde{t}_1^{*} \rightarrow gg) 
\frac{\widehat \alpha_S}{\pi}f(\epsilon)C_{\rm diagram}.
\eeq
Here, 
\beq
f(\epsilon) = \left( \frac{\pi \mu^2}{m^2_{\tilde{t}_1}} \right)^{\epsilon} 
\Gamma (1 + \epsilon),
\eeq
in keeping with the notation of ref. \cite{Hagiwara:1980nv}.  
Also, $C_F$ is the quadratic Casimir invariant, 
$C_A$ is the Casimir invariant of the adjoint representation, 
and $T_F$ is the index of the fundamental representation,
given for $SU(3)$ by $C_F = 4/3$, $C_A = 3$, and $T_F = 1/2$.   
We have combined diagrams $q1$ with $q2$ and $q3$ with $q4$ in the table 
because the individual self-energy diagrams are not proportional to the 
projector $g^{\mu \nu} - q^{\mu} q^{\nu} / q^2$. 
\begin{table}
\begin{tabular}[c]{|c|c|c|}
\hline
diagram & $C_{\rm diagram}$, three particle cut & $C_{\rm diagram}$, two 
particle cut \\
\hline
a & $C_A \left(\frac{5}{2} - \frac{\pi^2}{4}\right)$ & $0$ \\
\hline
b1 & $C_A \left(\frac{5}{2} - \frac{\pi^2}{3}\right)$ & $0$ \\
\hline
b2 & $C_A \left(-\frac{1}{8\epsIR^2} - \frac{5}{8\epsIR} 
- \frac{17}{8} + \frac{\pi^2}{12}\right)$ & $C_A \left(\frac{1}{8\epsIR^2} 
+ \frac{5}{8\epsIR} - \frac{3}{8\epsUV} + \frac{1}{4} \lntwo + \frac{1}{2} - \frac{\pi^2}{48}\right)$
\\
\hline
c1 & $0$ & $C_A \left( - \frac{3}{16\epsUV} - \frac{25}{48} 
- \frac{1}{24} \lntwo \right)$ 
\\
\hline
c2 & $0$ & $C_A \left( \frac{3}{2\epsUV} + \frac{7}{2} \right)$ \\
\hline
c3 & $0$ & $C_A \left( -\frac{3}{\epsUV} - \frac{11}{2} \right)$ \\
\hline
d1 & $C_A \left(\frac{1}{2\epsIR^2} + \frac{1}{\epsIR} - \frac{\pi^2}{6}\right)$ & $C_A \left( -\frac{1}{2\epsilon^2_{IR}} - \frac{1}{\epsIR} + \frac{3}{16\epsUV} - \frac{71}{48} + \frac{\pi^2}{3} + \frac{13}{24} \lntwo \right)$ \\
\hline
d2 & $C_A \left(-\frac{7}{4\epsIR^2} - \frac{3}{\epsIR} - \frac{23}{4} + \frac{7\pi^2}{6}\right)$ & $C_A \left( \frac{7}{4\epsilon^2_{IR}} + \frac{3}{\epsIR} - \frac{15}{8\epsUV} + \frac{11}{8} 
- \frac{7\pi^2}{6} - 2 \lntwo \right)$ \\
\hline
d3 & $C_A \left(\frac{9}{4\epsIR^2} + \frac{3}{\epsIR} + \frac{33}{4} - \frac{3\pi^2}{2} \right)$ & $C_A \left( -\frac{9}{4\epsilon^2_{IR}} - \frac{3}{\epsIR} + \frac{9}{2\epsUV} + \frac{9}{2} + \frac{3\pi^2}{2} \right)$ \\
\hline
e1 & $0$ & $C_A \left( 1 - \frac{\pi^2}{16} 
+ \frac{1}{2} \lntwo \right)$ \\
\hline
e2 & $C_A \left(\frac{1}{8\epsIR^2} + \frac{5}{8\epsIR} + \frac{17}{8} + \frac{\pi^2}{24}\right)$ & $C_A \left( - \frac{1}{8\epsilon^2_{IR}} - \frac{5}{8\epsIR} - \frac{21}{8} - \frac{11\pi^2}{48} + \frac{5}{4} \lntwo \right)$ \\
\hline
f1 & $0$ & $C_F \left( \frac{1}{4\epsUV} - \frac{1}{4} + \frac{5\pi^2}{16} 
- \frac{3}{2} \lntwo \right)$ \\
\hline
f2 & $0$ & $C_F \left( -\frac{1}{2\epsIR} - \frac{1}{\epsUV} - \frac{\pi^2}{2v}
- \frac{3}{2} 
- \frac{3\pi^2}{16}  - 4 \lntwo \right)$ \\
\hline
f3 & $0$ & $C_F \left( -\frac{1}{2\epsIR} - \frac{1}{4\epsUV} 
- \frac{\pi^2}{2v} - \frac{1}{4} 
 - \frac{3}{2} \lntwo \right)$ \\
\hline
f4 & $0$ & $C_F \left( \frac{2}{\epsIR} + \frac{1}{\epsUV} + \frac{2\pi^2}{v}
- \frac{1}{2} 
 + 6 \lntwo \right)$ \\
\hline
g1 & $0$ & $C_F \left( -\frac{1}{4\epsUV} + \frac{1}{4} - \frac{\pi^2}{16} - \frac{1}{2} \lntwo \right)$ \\
\hline
g2 & $0$ & $C_F \left( \frac{1}{\epsUV} + 2 - \frac{\pi^2}{16} 
+ 2 \lntwo \right)$ \\
\hline
g3 & $0$ & $C_F \left( \frac{1}{4\epsUV} + \frac{3}{4} + \frac{1}{2} \lntwo \right)$ \\
\hline
g4 & $0$ & $C_F \left( -\frac{1}{\epsUV} - \frac{5}{2} - 2 \lntwo \right)$ \\
\hline
h & $0$ & $\left( C_F - \frac{1}{2}C_A \right) \left( -\frac{1}{2\epsUV} - \frac{3}{2} \right)$ \\
\hline
i & $0$ & $\left( C_F - \frac{1}{4}C_A \right) \left( \frac{3}{4\epsUV} + \frac{3}{2} \lntwo + \frac{5}{2} \right)$ \\
\hline
j1 & $0$ & $\left( C_F - \frac{1}{4}C_A \right) \left( \frac{1}{\epsUV} + 3 \right)$ \\
\hline
j2 & $0$ & $\left( C_F - \frac{1}{4}C_A \right) \left( -\frac{4}{\epsUV} - 10 \right)$ \\
\hline
k & $0$ & $\left( C_F - \frac{1}{4}C_A \right) \left( \frac{3}{4\epsUV} + 2 + \frac{1}{2} \lntwo \right)$ \\
\hline
l1 & $0$ & $\left( C_F - \frac{1}{4}C_A \right) \left( -\frac{1}{2\epsUV} - \frac{7}{2} + \frac{\pi^2}{8} + \lntwo \right)$ \\
\hline
l2 & $0$ & $\left( C_F - \frac{1}{4}C_A \right) \left( \frac{2}{\epsUV} + 8 - \frac{3\pi^2}{8} - \lntwo \right)$ \\
\hline
\end{tabular}
\caption{Results for diagrams not involving propagator 
corrections to the tree level diagrams.\label{tab:nonpropagator}}
\end{table}
%%%%%%%%%%%%%%%%
\begin{table}
\begin{tabular}[c]{|c|c|c|}
\hline
diagram & $C_{\rm diagram}$, three particle cut & $C_{\rm diagram}$, two particle cut \\
\hline
m1 & $C_A \bigl( - \frac{5}{12\epsIR} - \frac{61}{36} \bigr) $ & $0$ \\
\hline
m2 & $0$ & $C_A \bigl( \frac{5}{6\epsIR} - \frac{5}{6\epsUV} \bigr) $ \\
\hline
m3 & $C_A \left( \frac{5}{4\epsIR} + \frac{17}{4} \right)$ & $C_A \left( \frac{5}{3\epsUV} - \frac{5}{3\epsIR} \right)$ \\
\hline
n1 & $0$ & $C_F \left(\frac{1}{2\epsIR} - \frac{1}{2\epsUV} \right)$ \\
\hline
n2 & $0$ & $C_F \left(\frac{1}{2\epsIR} - \frac{1}{2\epsUV} \right)$ \\
\hline
n3 & $0$ & $C_F \left(\frac{2}{\epsUV} - \frac{2}{\epsIR} \right)$ \\
\hline
o & $0$ & $C_F \left( -\frac{1}{2\epsUV} - \frac{3}{2} - \lntwo \right)$ \\
\hline
p1 & $ \sum_f T_F \Big( \frac{8}{9} + 
\frac{1}{3} \ln ( 
{m_f^2}/{4m_{\tilde{t}_1}^2} ) + 
h(m_f^2 / m_{\tilde{t}_1}^2) \Big)$ & 
$0$ 
\\
\hline
p2 & $0$ & $\sum_f T_F \left( \frac{2}{3\epsUV} + \frac{2}{3} - 
\frac{2}{3} \ln ({m_f^2}/{4m_{\tilde{t}_1}^2}) \right)$ \\
\hline
p3 & $ \sum_f T_F \Big( -\frac{8}{3} 
-\ln ( {m_f^2}/{4m_{\tilde{t}_1}^2}) 
- 3 h(m_f^2 / m_{\tilde{t}_1}^2) \Big) $ & $ 
\sum_f T_F \left( -\frac{4}{3\epsUV} - \frac{2}{3} + 
\frac{4}{3} \ln ( {m_f^2}/{4m_{\tilde{t}_1}^2}) \right)$ \\
\hline
q1 + q2 & $0$ & $ T_F \left( 
\frac{1}{6\epsUV} + \frac{1}{6} + \frac{1}{3} \lntwo \right)$\\
\hline
q3 + q4 & $0$ & $ T_F \left( 
-\frac{1}{3\epsUV} - \frac{1}{6} - \frac{2}{3} \lntwo \right)$\\
\hline
\end{tabular}
\caption{Results for diagrams involving propagator corrections to 
the tree-level diagrams. The function $h(m_f^2 / m_{\tilde{t}_1}^2)$ 
is defined in eq.~(\ref{eq:h(r)}).
\label{tab:propagator}}
\end{table}
%%%%%%%%%%%%%%%%

Taking the sum of the diagrams, we find the next-to-leading order result
\beq
v \sigma^{(1)} \left( \tilde{t}_1 \tilde{t}_1^{*} \rightarrow gg \right) & = & 
v \sigma^{(0)} \left( \tilde{t}_1 \tilde{t}_1^{*} \rightarrow gg \right) 
\bigg\{ 1 + f(\epsilon) \frac{\widehat \alpha_S}{\pi} \bigg[ \frac{b_0}{2 \epsUV} + 
\left( \frac{199}{18} - \frac{13 \pi^2}{24} \right) C_A 
\nonumber \\
 & & + \left( \frac{\pi^2}{v} -\frac{7}{2} - \frac{\pi^2}{8} +  
 \left( \frac{1}{2} - \frac{\pi^2}{8} \right) \delta \right) C_F 
\nonumber \\
 & & + \left( -\frac{16}{9}(n_{\rm light} + n_t) 
       - 2 n_t h(m_t^2/m_{\tilde{t}_1}^2) -
     \frac{1}{3} \lntwo \right) T_F \bigg] \bigg\},
\eeq
where $\delta$ is either 1 or 0 depending on whether or not the 
four-point squark interaction in Figure \ref{fig:feynmanrules} is 
included,\footnote{In the MSSM,
$\delta = 1$. However, one can 
imagine
non-supersymmetric theories with fundamental strongly interacting
scalars, in which these formulas would apply with $\delta=0$.} 
$n_{\rm light} = 5$ is the number of light quarks, and
$n_t =1$ or 0 depending on whether or not the top quark is included in 
the effective theory. In this formula, we have written
\beq
b_0 & = & \frac{11}{3} C_A - \frac{4}{3} T_F (n_{\rm light} + n_t) 
- \frac{1}{3} T_F \\
 & & \nonumber \\
h \left( r \right) & = & \frac{2}{9} (4 - r) \sqrt{1-r} - \frac{8}{9}  
- \frac{2}{3} \ln(1+\sqrt{1-r})
+ \frac{2}{3} \lntwo 
\label{eq:h(r)} \\
 & & \nonumber \\
r & = & {m_t^2}/{m_{\tilde{t}_1}^2}.
\eeq
The function $h(r)$ is defined so that it parametrizes the effects of 
a non-zero top-quark mass.  
In the limit that the top quark is massless compared to the top squark 
we have $h(0) = 0$, and when the masses are identical we have 
$h(1) = -\frac{8}{9} + \frac{2}{3} \lntwo$.

The one-loop order correction to the diphoton cross-section can now be 
found simply by dropping the diagrams that involve gluon self-coupling or 
real gluon emission (equivalent to setting $C_A = 0$) as well as any 
vacuum polarization diagrams (which no longer involve strong couplings), 
then making the replacement 
$\hat g_3 T^{aj}_k \rightarrow Qe\delta^j_k$ at 
vertices to change gluons into photons. Following this procedure, we find
\beq
v \sigma^{(1)} \left( \tilde{t}_1 \tilde{t}_1^{*} 
\rightarrow \gamma \gamma \right) & = & 
v \sigma^{(0)} \left( \tilde{t}_1 \tilde{t}_1^{*} 
\rightarrow \gamma \gamma \right) 
\bigg\{ 1 + f(\epsilon) \frac{\widehat \alpha_S}{\pi} C_F
\bigg[ \frac{\pi^2}{v} -\frac{11}{2} + \frac{\pi^2}{8} 
- 2 \lntwo 
\nonumber \\ && 
+  \left( \frac{1}{2} - \frac{\pi^2}{8} \right) \delta \bigg] \bigg\}.
\label{eq:gammagammaNR}
\eeq

In the \MSbar renormalization scheme, the bare coupling 
$\widehat \alpha_S$ is written in terms 
of the renormalized running coupling $\alpha_S(Q)$ using
\beq
\widehat \alpha_S = \alpha_S \bigg[ 1 - \frac{\alpha_S}{4 \pi} b_0 \left(\frac{1}{\epsUV} + 
\ln \left( {4\pi \mu^2}/{Q^2} \right) - \gamma_E \right) \bigg].
\eeq
The gluon cross section as a function of the 
renormalized \MSbar coupling $\alpha_S$ and the renormalization scale 
$Q$ can therefore be written as
\beq
v \sigma^{(1)} \left( \tilde{t}_1 \tilde{t}_1^{*} \rightarrow gg \right) & = & 
v \sigma^{(0)} \left( \tilde{t}_1 \tilde{t}_1^{*} \rightarrow gg \right) 
\bigg\{ 1 + \frac{\alpha_S}{\pi} \bigg[ \frac{b_0}{2} 
\ln \biggl( \frac{Q^2}{4m_{\tilde{t}_1}^2} \biggr) + 
\left( \frac{199}{18} - \frac{13 \pi^2}{24} \right) C_A 
\nonumber \\
 & & + \left( \frac{\pi^2}{v} -\frac{7}{2} - \frac{\pi^2}{8} 
 + \left( \frac{1}{2} - \frac{\pi^2}{8} \right) \delta \right) C_F 
\nonumber \\
 & & + \left( -\frac{16}{9}(n_{\rm light} + n_t) 
- 2 n_t h(m_t^2 / m_{\tilde{t}_1}^2) 
- \frac{1}{3} \lntwo \right) T_F \bigg] \bigg\}.
\eeq
For the diphoton final state, one can simply replace the bare coupling 
$\widehat \alpha_S$
by the \MSbar coupling $\alpha_S$ in eq.~(\ref{eq:gammagammaNR}) 
to obtain the corresponding
renormalized result for $v \sigma^{(1)} \left( \tilde{t}_1 \tilde{t}_1^{*} \rightarrow \gamma\gamma \right)$, since the QCD coupling does not appear in
the tree-level result in that case.

The ratio of partial widths is now obtained at 
next-to-leading order by
\beq
R^{(1)} \equiv 
\frac{\Gamma^{(1)}(\eta_{\tilde{t}} \rightarrow 
\gamma \gamma)}{\Gamma^{(1)}(\eta_{\tilde{t}} \rightarrow 
\mbox{hadrons})}.
\eeq
(Here we write ``hadrons" to subsume the $gg$, $ggg$, and the partial $gq\overline q$ parton-level contributions.)
The $1/v$ Coulomb singularity does not appear in this ratio, since it can
be absorbed into a redefinition of the bound-state wavefunction factor
\cite{ref:coulsing}, at least at order $\alpha_S$ in the approximation of
a Coulombic bound state, and the redefined bound-state factor in turn
cancels from the ratio of decay rates. This redefinition simply removes
the $1/v$ part
as it is expanded to next-to-leading order in $\alpha_S$. There may
remain some small residual dependence on $1/v$ proportional to
$\alpha_S^2$, due to the fact that the bound-state potential is actually
not exactly Coulombic, but this is beyond the scope of the present work
since we work only to order $\alpha_S$ in the ratio.
Our final result is
\beq
R^{(1)} & = & 
\frac{8\alpha^2}{9\alpha_S^2} 
\bigg\{ 1 + \frac{\alpha_S}{\pi} \bigg[ -\frac{b_0}{2} 
\ln \biggl( \frac{Q^2}{4m_{\tilde{t}_1}^2} \biggr) + 
\left(\frac{13 \pi^2}{24}-\frac{199}{18} \right) C_A  
 + \left(\frac{\pi^2}{4} -2 - 2 \lntwo \right) C_F \nonumber \\
 & & + \left( \frac{16}{9}(n_{\rm light} + n_t) 
+ 2 n_t h(m_t^2 / m_{\tilde{t}_1}^2) 
+ \frac{1}{3} \lntwo \right) T_F \bigg] \bigg\}.
\label{eq:Roneresult}
\eeq
An interesting feature of this result is that 
the term proportional to $C_A$ is the 
same for the stoponium and the corresponding quarkonium 
calculation
as can be seen by
comparing\footnote{Note that $R^{(1)}$ 
in ref.~\cite{Hagiwara:1980nv} is the reciprocal of our definition.}
eq.~(\ref{eq:Roneresult}) to eqs.~(4.1)-(4.4) of 
ref.~\cite{Hagiwara:1980nv}. 

%%%%%%%%%%%%%%%%%%%%%%%%%%%%%%%%%%%%%%%%%%%%%%%%%%%%%%%%%%%%%%%%%%%%%%%%%%%%%%%%
\section{Decays to Higgs scalar bosons\label{sec:hh}}
\setcounter{footnote}{1}
\setcounter{equation}{0}

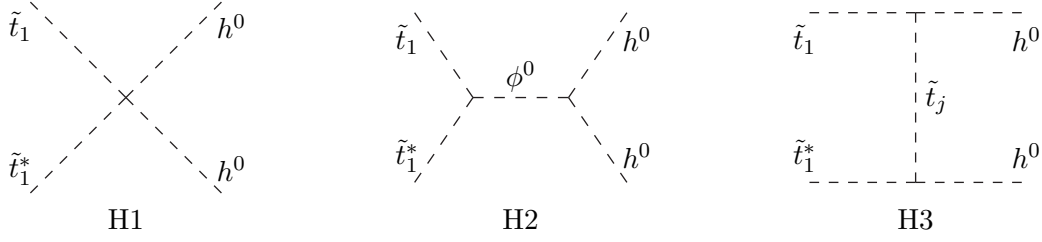
\begin{figure}[!t]
\begin{center}
\begin{picture}(80,80)(-40,-40)
\DashLine(36,36)(0,0){4}
\DashLine(36,-36)(0,0){4}
\DashLine(-36,36)(0,0){4}
\DashLine(-36,-36)(0,0){4}
\Text(-40,27)[]{$\tilde{t}_1$}
\Text(-40,-27)[]{$\tilde{t}_1^*$}
\Text(40,27)[]{$h^0$}
\Text(40,-27)[]{$h^0$}
\Text(0,-46)[c]{H1}
\end{picture}
~~~~~~~~~~~~~~~~~
\begin{picture}(80,80)(-40,-40)
\DashLine(40,32)(18,0){4}
\DashLine(40,-32)(18,0){4}
\DashLine(-40,32)(-18,0){4}
\DashLine(-40,-32)(-18,0){4}
\DashLine(18,0)(-18,0){4}
\Text(-43,22)[]{$\tilde{t}_1$}
\Text(-43,-22)[]{$\tilde{t}_1^*$}
\Text(44,22)[]{$h^0$}
\Text(44,-22)[]{$h^0$}
\Text(0,8)[c]{$\phi^0$}
\Text(0,-46)[c]{H2}
\end{picture}
~~~~~~~~~~~~~~~~~
\begin{picture}(80,80)(-40,-40)
\DashLine(40,32)(0,32){4}
\DashLine(40,-32)(0,-32){4}
\DashLine(-40,32)(0,32){4}
\DashLine(-40,-32)(0,-32){4}
\DashLine(0,32)(0,-32){4}
\Text(-42,22)[]{$\tilde{t}_1$}
\Text(-42,-22)[]{$\tilde{t}_1^*$}
\Text(42,22)[]{$h^0$}
\Text(42,-22)[]{$h^0$}
\Text(8,0)[c]{$\tilde{t}_j$}
\Text(0,-46)[c]{H3}
\end{picture}
\end{center}
\caption{\label{fig:higgstree}
The tree-level diagrams for the annihilation of 
$\tilde{t}_1\tilde{t}_1^*$ into $h^0h^0$.}  
\end{figure}

We now consider the one-loop radiative corrections to $\eta_{\tilde t} 
\rightarrow
h^0 h^0$, where $h^0$ is the lightest Higgs scalar boson in supersymmetry.
In this case we calculate the partial widths directly
rather than using the cut 
method.
The tree-level diagrams contributing to this annihilation decay are shown in
figure~\ref{fig:higgstree}.
The corresponding annihilation cross-section in the $v \rightarrow 0$ limit
can be written as \cite{Drees:1993uw}
(see also \cite{Barger:1988sp,Martin:2008sv}):
\beq	
v \sigma^{(0)} (\tilde{t}_1\tilde{t}_1^{*} \rightarrow h^0h^0) &=& 
\frac{N_c}{64\pi m_{\tilde{t}_1}^2}  
(1- m_{h^0}^2/m_{\tilde t_1}^2)^{1/2}
\bigl ( \lambda_1 + \lambda_2 + \lambda_3 \bigr )^2,
\label{eq:LOhhresult}
\eeq
where, in the notation of ref.~\cite{SPMcouplingsnotation}, the effective 
couplings are:
\beq
\lambda_1 &=& \lambda_{h^0 h^0 \tilde t_1 \tilde t_1^*},
\label{eq:deflambdaone}
\\
\lambda_2 &=& \sum_{\phi^0 = h^0,H^0} 
%\frac{
{\lambda_{\phi^0\tilde t_1 \tilde t_1^*} 
 \lambda_{\phi^0 h^0 h^0}
}/{(4 m^2_{\tilde t_1} - m^2_{\phi^0})}
,
\label{eq:deflambdatwo}
\\
\lambda_3 &=& \sum_{j=1,2} 
-2 | \lambda_{h^0 \tilde t_1 \tilde t_j^*}|^2/(
m^2_{\tilde t_1} + m^2_{\tilde t_j} - m^2_{h^0}).
\label{eq:deflambdathree}
\eeq
Using eq.~\ref{eq:Gamgamgam}),
the tree-level ratio of $\gamma\gamma$ to $h^0 h^0$ partial widths is
therefore
\beq
\Gamma^{(0)}(\gamma\gamma)/\Gamma^{(0)}(h^0 h^0)
= 
\frac{2048 \pi^2 \alpha^2}{81 (\lambda_1 + \lambda_2 + \lambda_3)^2}
(1- m_{h^0}^2/m_{\tilde t_1}^2)^{-1/2}.
\eeq

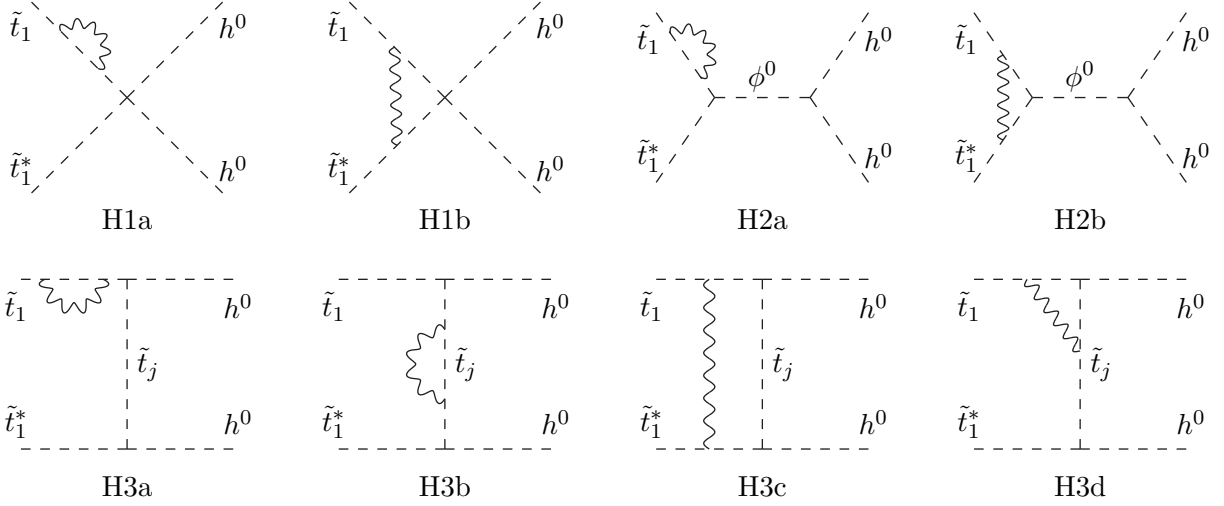
\begin{figure}[!t]
\begin{center}
\begin{picture}(80,80)(-40,-40)
\DashLine(36,36)(0,0){4}
\DashLine(36,-36)(0,0){4}
\DashLine(-36,36)(0,0){4}
\DashLine(-36,-36)(0,0){4}
\PhotonArc(-18,18)(10,-45,135){1.8}{5.5}
\Text(-40,27)[]{$\tilde{t}_1$}
\Text(-40,-27)[]{$\tilde{t}_1^*$}
\Text(40,27)[]{$h^0$}
\Text(40,-27)[]{$h^0$}
\Text(0,-46)[c]{H1a}
\end{picture}
~~~~~~~~~
\begin{picture}(80,80)(-40,-40)
\DashLine(36,36)(0,0){4}
\DashLine(36,-36)(0,0){4}
\DashLine(-36,36)(0,0){4}
\DashLine(-36,-36)(0,0){4}
\Photon(-19,19)(-18,-18){-2.0}{5.5}
\Text(-40,27)[]{$\tilde{t}_1$}
\Text(-40,-27)[]{$\tilde{t}_1^*$}
\Text(40,27)[]{$h^0$}
\Text(40,-27)[]{$h^0$}
\Text(0,-46)[c]{H1b}
\end{picture}
~~~~~~~~~
\begin{picture}(80,80)(-40,-40)
\DashLine(40,32)(18,0){4}
\DashLine(40,-32)(18,0){4}
\DashLine(-40,32)(-18,0){4}
\DashLine(-40,-32)(-18,0){4}
\DashLine(18,0)(-18,0){4}
\PhotonArc(-29,16)(10,-55.4915,124.509){1.8}{5.5}
\Text(-43,22)[]{$\tilde{t}_1$}
\Text(-43,-22)[]{$\tilde{t}_1^*$}
\Text(44,22)[]{$h^0$}
\Text(44,-22)[]{$h^0$}
\Text(0,8)[c]{$\phi^0$}
\Text(0,-46)[c]{H2a}
\end{picture}
~~~~~~~~~
\begin{picture}(80,80)(-40,-40)
\DashLine(40,32)(18,0){4}
\DashLine(40,-32)(18,0){4}
\DashLine(-40,32)(-18,0){4}
\DashLine(-40,-32)(-18,0){4}
\DashLine(18,0)(-18,0){4}
\Photon(-29,16)(-29,-16){-2.1}{5.5}
\Text(-43,22)[]{$\tilde{t}_1$}
\Text(-43,-22)[]{$\tilde{t}_1^*$}
\Text(44,22)[]{$h^0$}
\Text(44,-22)[]{$h^0$}
\Text(0,8)[c]{$\phi^0$}
\Text(0,-46)[c]{H2b}
\end{picture}

\vspace{0.7cm}

\begin{picture}(80,80)(-40,-40)
\DashLine(40,32)(0,32){4}
\DashLine(40,-32)(0,-32){4}
\DashLine(-40,32)(0,32){4}
\DashLine(-40,-32)(0,-32){4}
\DashLine(0,32)(0,-32){4}
\PhotonArc(-20,32)(11,-180,0){2.2}{5.5}
\Text(-42,22)[]{$\tilde{t}_1$}
\Text(-42,-22)[]{$\tilde{t}_1^*$}
\Text(42,22)[]{$h^0$}
\Text(42,-22)[]{$h^0$}
\Text(8,0)[c]{$\tilde{t}_j$}
\Text(0,-46)[c]{H3a}
\end{picture}
~~~~~~~~~
\begin{picture}(80,80)(-40,-40)
\DashLine(40,32)(0,32){4}
\DashLine(40,-32)(0,-32){4}
\DashLine(-40,32)(0,32){4}
\DashLine(-40,-32)(0,-32){4}
\DashLine(0,32)(0,-32){4}
\PhotonArc(0,0)(13,90,270){2.0}{5.5}
\Text(-42,22)[]{$\tilde{t}_1$}
\Text(-42,-22)[]{$\tilde{t}_1^*$}
\Text(42,22)[]{$h^0$}
\Text(42,-22)[]{$h^0$}
\Text(8,0)[c]{$\tilde{t}_j$}
\Text(0,-46)[c]{H3b}
\end{picture}
~~~~~~~~~
\begin{picture}(80,80)(-40,-40)
\DashLine(40,32)(0,32){4}
\DashLine(40,-32)(0,-32){4}
\DashLine(-40,32)(0,32){4}
\DashLine(-40,-32)(0,-32){4}
\DashLine(0,32)(0,-32){4}
\Photon(-20,32)(-20,-32){-2.1}{7.5}
\Text(-42,22)[]{$\tilde{t}_1$}
\Text(-42,-22)[]{$\tilde{t}_1^*$}
\Text(42,22)[]{$h^0$}
\Text(42,-22)[]{$h^0$}
\Text(8,0)[c]{$\tilde{t}_j$}
\Text(0,-46)[c]{H3c}
\end{picture}
~~~~~~~~~
\begin{picture}(80,80)(-40,-40)
\DashLine(40,32)(0,32){4}
\DashLine(40,-32)(0,-32){4}
\DashLine(-40,32)(0,32){4}
\DashLine(-40,-32)(0,-32){4}
\DashLine(0,32)(0,-32){4}
\Photon(-20,32)(0,5){-2.2}{5.5}
\Text(-42,22)[]{$\tilde{t}_1$}
\Text(-42,-22)[]{$\tilde{t}_1^*$}
\Text(42,22)[]{$h^0$}
\Text(42,-22)[]{$h^0$}
\Text(8,0)[c]{$\tilde{t}_j$}
\Text(0,-46)[c]{H3d}
\end{picture}
\end{center}
\caption{\label{fig:higgsNLO}
The one-loop QCD diagrams for the annihilation of 
$\tilde{t}_1\tilde{t}^*_1$ into $h^0h^0$.}  
\end{figure}

The one-loop QCD corrections to the stoponium decay to $h^0h^0$ 
are due to the diagrams shown in
figure \ref{fig:higgsNLO}.
Note that there are no gluon emission diagrams to this process at this order, 
because the initial and final states are both
color singlets.
In this paper, we will 
neglect\footnote{The decay of stoponium to $h^0h^0$ is especially 
important in electroweak-scale baryogenesis models \cite{baryo}, 
\cite{baryoDM}, \cite{baryonew} 
that require light $\tilde t_1$. In those models, $m_{\tilde t_2}$ is 
necessarily very large, making this approximation extremely good.} in 
$\lambda_3$ the contribution of the heavier top-squark mass 
eigenstate $\tilde t_2$, which would otherwise entail a somewhat more 
complicated kinematic loop integration.
At one-loop order in QCD and in the limit of small 
$v$, the corresponding $\tilde t_1 \tilde t_1^* \rightarrow
h^0h^0$ result is:
\beq	
v \sigma^{(1)} (\tilde{t}_1\tilde{t}_1^{*} \rightarrow h^0h^0) &=& 
\frac{N_c }{64\pi m_{\tilde{t}_1}^2}  (1- m_{h^0}^2/m_{\tilde t_1}^2)^{1/2}
\left [ C (\widehat \lambda_1 + \widehat \lambda_2) + 
C' \widehat \lambda_3  \right ]^2,
\eeq
where, 
in terms of the individual diagram contributions 
given in Table \ref{table:hh},%
%%%%%%%%%%%%%%%%%%%%%
\begin{table}
\begin{tabular}[c]{|c|c|}
\hline
diagram & $C_{\rm diagram}$ \\
\hline
\phantom{x} H1a, H2a, H3a \phantom{x} & $
\frac{1}{2\epsUV} - \frac{1}{2\epsIR} 
$ \\[4pt]
\hline
H1b, H2b & $
\frac{1}{2\epsIR} + \frac{1}{4 \epsUV} 
+ \frac{\pi^2}{2v} 
- \frac{1}{2} 
+ \frac{3}{2} \lntwo 
$ \\[4pt]
\hline
H3b & $
\frac{1}{2\epsUV} + 1 + \lntwo + \frac{h}{2(1-h)}  \ln(2- h) 
$ \\[4pt]
\hline
H3c & $
\frac{1}{2\epsIR} + \frac{\pi^2}{2v}
-1 - \ln(1-h/2) -\frac{h}{2-h} k_1(h)- \frac{1}{2} k_2(h) 
$ \\[4pt]
\hline
H3d & $
\phantom{xx}
\frac{1}{2\epsUV} + 1 + \lntwo - \frac{2-h}{2(1-h)} \ln(2- h) 
+\frac{1}{2} k_1(h) + k_2(h) 
$ 
\phantom{xx}
\\[4pt]
\hline
\end{tabular}
\caption{Results for one-loop radiative corrections to 
$\tilde t_1 \tilde t_1^* \rightarrow h^0 h^0$, corresponding 
to the diagrams in figure \ref{fig:higgsNLO}
and appearing in eqs.~(\ref{eq:defChat}), (\ref{eq:defChatprime}).
\label{table:hh}}
\end{table}
%%%%%%%%%%%%%%%%%%%%%%%%%%%
\beq
C &=& 1 + C_F \frac{\alpha_S}{\pi} f(\epsilon) 
\left (C_{H1a} + C_{H1b} \right ),
\label{eq:defChat}
\\
C'&=& 1 + C_F \frac{\alpha_S}{\pi} f(\epsilon) 
\left (C_{H3a} + C_{H3b} + C_{H3c} + C_{H3d} \right ).
\label{eq:defChatprime}
\eeq
The couplings $\widehat \lambda_{1,2,3}$ are given by the same
formulas as eq.~(\ref{eq:deflambdaone})-(\ref{eq:deflambdathree}), but with
bare couplings, denoted by
$\widehat \lambda_{h^0 h^0 \tilde t_1 \tilde t_1^*}$, 
$\widehat \lambda_{\phi^0\tilde t_1 \tilde t_j^*}$, and 
$\widehat \lambda_{\phi^0 h^0 h^0}$,
in place of their unhatted counterparts. Also,
\beq
h \equiv {m_{h^0}^2}/{m_{\tilde{t}_1}^2},
\eeq
and we have defined functions
\beq
k_1(h) &=& B \tan^{-1}(h B/(2-h)),
\\
k_2(h) &=& \frac{2-h}{2A} {\rm Re}\Bigl [
\dilog \Bigl (\frac{1+A}{1+ i AB} \Bigr ) 
-\dilog \Bigl (\frac{1-A}{1+ i AB} \Bigr ) 
\nonumber \\ &&
+ \frac{1}{2} \dilog \Bigl (1 - \frac{2A}{2-h} \Bigr )
- \frac{1}{2} \dilog \Bigl (1 + \frac{2A}{2-h} \Bigr )
\Bigr ],
\phantom{xxxxx}
\eeq
with $A = \sqrt{1-h}$ and $B = \sqrt{4/h -1}$, and Li${}_2(x)$ is the
dilogarithm function (also known as the Spence function).
These functions have values  $k_1(0) = 2$ and 
and $k_2(0) = -\pi^2/8$ for the extreme limit 
$m_{h^0} \ll m_{\tilde t_1}$, and
$k_1(1) = \pi/\sqrt{3}$
and $k_2(1) = -\pi/2\sqrt{3}$ near threshold for the decay.
The calculation of these diagrams relies on loop integrals that 
can be found from
refs.~\cite{tHooft:1978xw,Beenakker:1988bq} 
by taking the $v\rightarrow 0$ limit with appropriate
special cases of momenta and masses.
Note that the $1/\epsIR$ poles cancel in eqs.~(\ref{eq:defChat})
and (\ref{eq:defChatprime}), as required.

The bare couplings can be written in terms of the renormalized 
running
\MSbar scheme couplings $\lambda_{h^0 h^0 \tilde t_1 \tilde t_1^*}$,
$\lambda_{\phi^0 \tilde t_1 \tilde t_1^*}$, and 
$\lambda_{\phi^0 h^0 h^0}$, at one-loop order in QCD, as:
\beq
\widehat{\lambda}_{h^0 h^0 \tilde t_1 \tilde t_1^*} & = & 
\lambda_{h^0 h^0 \tilde t_1 \tilde t_1^*} 
\biggl [ 1 - 
\frac{3\alpha_S}{4 \pi} \Bigl(\frac{1}{\epsUV} + 
\ln \left( {4\pi \mu^2}/{Q^2} \right) - \gamma_E \Bigr) \biggr ]  ,
\\
\widehat{\lambda}_{\phi^0 \tilde t_1 \tilde t_1^*} & = & 
\lambda_{\phi^0 \tilde t_1 \tilde t_1^*} \biggl [ 1 - 
\frac{3\alpha_S}{4 \pi} \Bigl (\frac{1}{\epsUV} + 
\ln \left( {4\pi \mu^2}/{Q^2} \right) - \gamma_E \Bigr) \biggr ] ,
\\
\widehat \lambda_{\phi^0 h^0 h^0} &=& \lambda_{\phi^0 h^0 h^0},
\eeq
which eliminates the $1/\epsilon_{\rm UV}$ dependence of the result
up to terms of ${\cal O}(\alpha_S^2)$.
It follows that
\beq	
v \sigma^{(1)} (\tilde{t}_1\tilde{t}_1^{*} \rightarrow h^0h^0) &=& 
\frac{N_c}{64\pi m_{\tilde{t}_1}^2}  (1- m_{h^0}^2/m_{\tilde t_1}^2)^{1/2}
K \left ( \lambda_1 + \lambda_2 + K' \lambda_3 \right )^2,
\label{eq:NLOhhresult}
\eeq
where $\lambda_{1,2,3}$ are as given in 
eqs.~(\ref{eq:deflambdaone})-(\ref{eq:deflambdathree}), with
$\lambda_{h^0 h^0 \tilde t_1 \tilde t_1^*}$,
$\lambda_{\phi^0 \tilde t_1 \tilde t_1^*} $, and
$\lambda_{\phi^0 h^0 h^0}$ taken to be \MSbar 
couplings, the masses renormalized on-shell,
and
\beq
K &=& 1 + C_F \frac{\alpha_S}{\pi} \left [
\frac{\pi^2}{v} + \frac{3}{2} \ln (Q^2/4 m_{\tilde t_1}^2) 
-1 + 3 \lntwo \right ]
,
\label{eq:defK}
\\
K'&=& 1 + C_F \frac{\alpha_S}{\pi} \biggl [
\frac{3}{4} \ln (Q^2/4 m_{\tilde t_1}^2) +
%% version 2 correction: 1/2 -> 3/2 in the following line
\frac{3}{2} - \frac{1}{2} \lntwo - 2 \ln(1-h/2) 
\nonumber \\ &&
+ \Bigl (\frac{1}{2} - \frac{h}{2-h} \Bigr ) k_1(h)
+ \frac{1}{2} k_2(h)
\biggr ].
\label{eq:defKprime}
\eeq
By comparing eqs.~(\ref{eq:gammagammaNR}), 
(\ref{eq:LOhhresult}), 
and
(\ref{eq:NLOhhresult}),
we obtain
\beq
\frac{\Gamma^{(1)}(\gamma\gamma)}{\Gamma^{(1)}(h^0h^0)}
&=&
\frac{\Gamma^{(0)}(\gamma\gamma)}{\Gamma^{(0)}(h^0h^0)}
\frac{(\lambda_1 + \lambda_2 + \lambda_3)^2}
     {(\lambda_1 + \lambda_2 + K'\lambda_3)^2}
\Bigl ( 1 - C_F \frac{\alpha_S}{\pi} \Bigl [
\frac{3}{2} \ln (Q^2/4 m_{\tilde t_1}^2) +4 + 5 \lntwo
\Bigr ] \Bigr ),\phantom{xxx}
\label{eq:gammagammaoverhh}
\eeq
written in terms of \MSbar couplings and on-shell masses.

%%%%%%%%%%%%%%%%%%%%%%%%%%%%%%%%%%%%%%%%%%%%%%%%%%%%%%%%%%%%%%%%%%%%%%%%%%%%%%%%
\section{Numerical results\label{sec:numerical}}
\setcounter{footnote}{1}
\setcounter{equation}{0}

In this section, we will examine the numerical impact of the radiative 
corrections found above in typical realistic cases. 
In many models, the 
dominant final state is hadrons coming from the $gg$ or $ggg$ 
parton-level process at next-to-leading order in QCD. As found in 
ref.~\cite{Martin:2008sv}, this often amounts to 90\% or more of the 
total decay width. 
(In particular, even when it is kinematically 
allowed, the $t\overline t$ final state is typically dominated by the 
$gg$ final state and has a branching ratio of only a few per cent, 
unless there is a resonant Higgs exchange contribution.)
Therefore, for simplicity we will begin by considering 
the idealized case that only diphoton and gluon-induced hadronic final 
states are included.

The relevant QCD group theory invariants for $SU(3)_c$ are 
$C_A = N_c = 3$, $C_F = (N_c^2 - 1)/2 N_c = 4/3$,
$T_F = 1/2$, and $n_{\rm light} = 5$. 
%% New in version 3:
At two-loop order, the QCD coupling runs according to
\cite{Srednicki:2007qs}
\beq
Q \frac{d\alpha_S}{dQ} &=& -\frac{b_0}{2 \pi} \alpha_S^2 
- \frac{b_1}{4 \pi^2} \alpha_S^3 ,
\label{eq:runningalphaS}
\\
b_0  &=&  \frac{23}{3} - \frac{2}{3} n_t - \frac{1}{6} n_s ,
\\
b_1  &=&  \frac{58}{3} - \frac{19}{3} n_t - \frac{11}{6} n_s .
\eeq
where $n_t$ and $n_s$ are either 0 or 1 depending on whether or not the 
top and stop, respectively, are included in the effective theory.  
Although we have not included the QED correction diagrams 
to the annihilation process, 
we do incorporate the running of the QED coupling according to
\beq
\alpha(Q)  =  
\frac{\alpha(Q_0)}{1 + \left( b^{EM}_0 \alpha(Q_0)/2 \pi \right) \ln (Q/Q_0)}
,
\qquad
b^{EM}_0  =  -\frac{80}{9} - \frac{16}{9} n_t - \frac{4}{9} n_s .
\label{eq:runningalpha}
\eeq
We take as inputs $M_Z = 91.18$ GeV, $m_t = 172.5$ GeV, and
$\alpha_S^{(5)}(M_Z) = 0.118$, 
$\alpha^{(5)}(M_Z) = 1/128.0$ as the
running \MSbar coupling inputs in the 5-quark Standard Model.
We then determine \MSbar couplings 
$\alpha_S(m_{\tilde t_1})$ and $\alpha(m_{\tilde t_1})$ by running
below $Q = m_t$  using the 5-quark two-loop renormalization group equations,
and, if $m_{\tilde t_1} > m_t$, 
running above $Q=m_t$ using the 6-quark renormalization group equations. 

Having determined the input parameter values at $Q = m_{\tilde t_1}$, we 
then match\footnote{We neglect threshold corrections when 
matching between different effective field theories, because they 
are of relative order $\alpha_S^2$ and 
numerically small compared to other sources of error and uncertainty.} 
onto and work in the effective theory in which the lighter top squark is 
always included (even for $Q < m_{\tilde t_1}$), so that $\alpha_S(Q)$ 
and $\alpha(Q)$ are obtained by running the renormalization group 
equations eq.~(\ref{eq:runningalphaS}) and (\ref{eq:runningalpha}) with 
$n_s=1$ and either $n_t=1$ (if $m_{\tilde t_1} > m_t$) or $n_t = 0$ (if 
$m_{\tilde t_1} < m_t$). We then evaluate the ratios of decay widths of 
stoponium into $\gamma\gamma+X$ and into hadrons at tree-level and at 
next-to-leading order, obtained from eqs.~(\ref{eq:Rzero}) and 
(\ref{eq:Roneresult}), respectively, self-consistently using the same 
value for $n_t$.

%%%%%%%%%%%%%%%%%%%%%%%%%%%%%%%%%%%%%%%%%%%%%%%%%%%%%%%%%%%%%%%%%%%%%%%%%%%%%
\begin{figure}[!tp]
\includegraphics[width=7.5cm,angle=0]{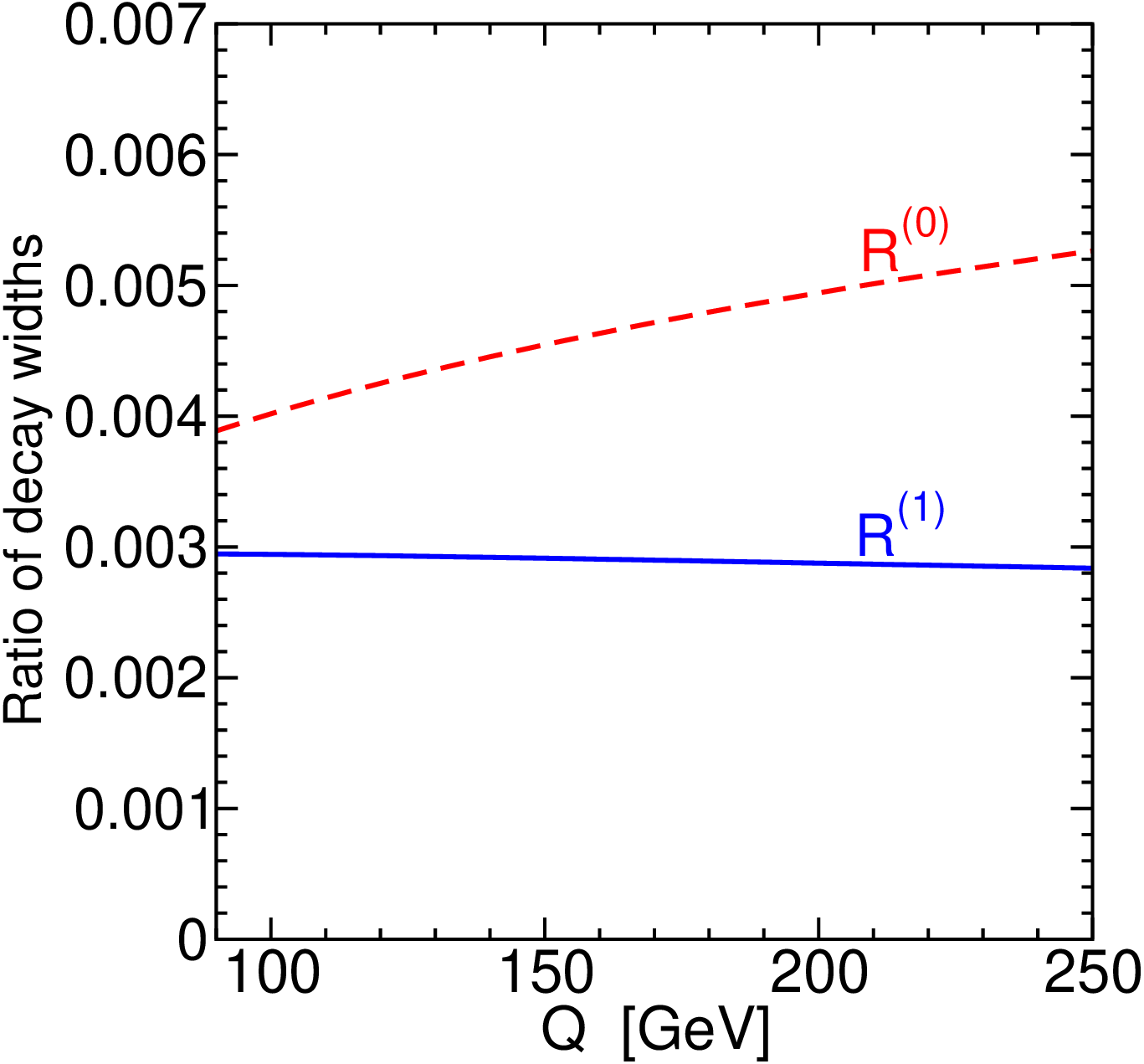}~~
\includegraphics[width=7.5cm,angle=0]{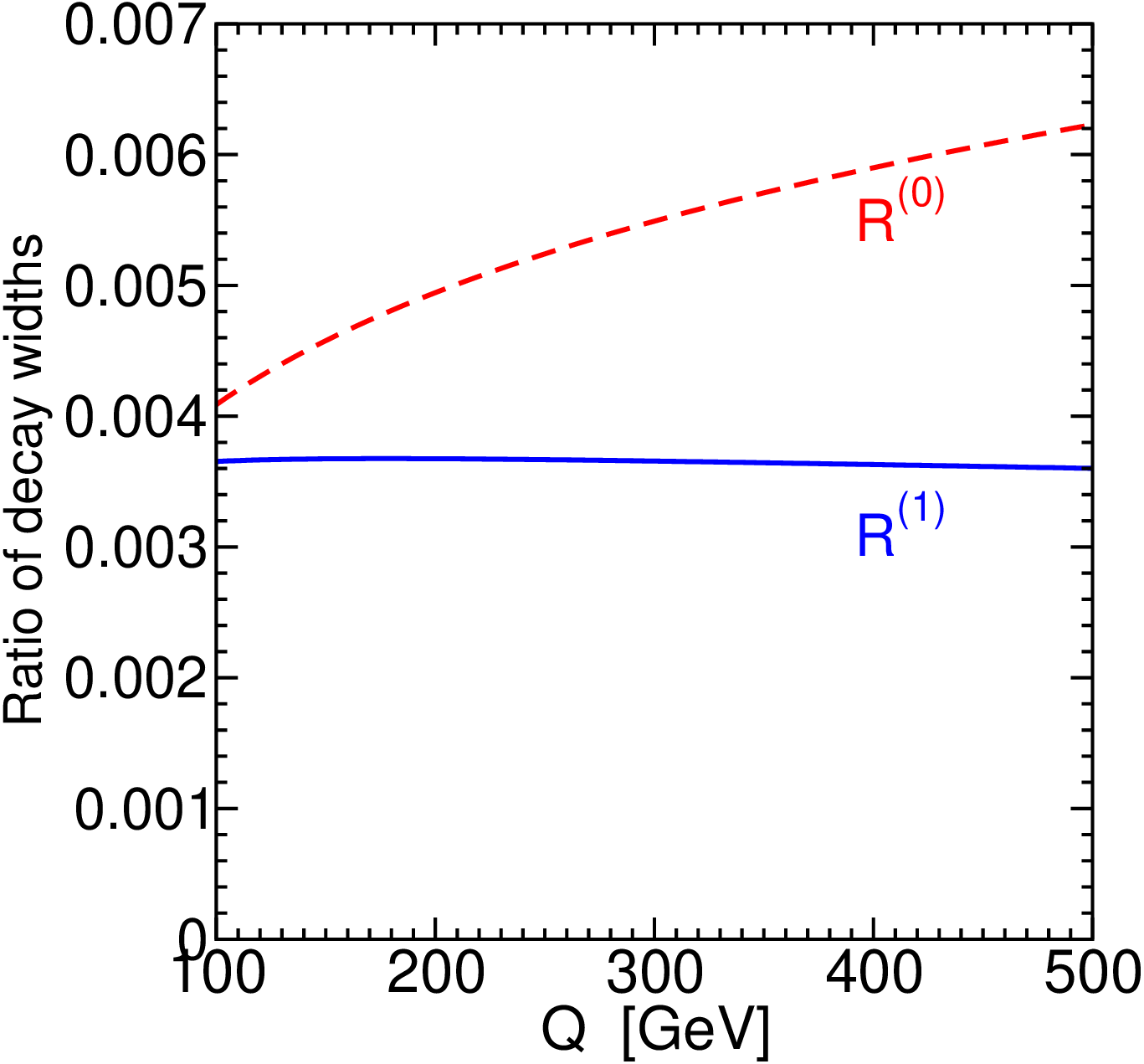}
\caption{\label{fig:RvsQ}
The dependence of the ratio of decay widths of stoponium
into $\gamma\gamma+X$ and into hadrons is shown at tree level
($R^{(0)}$, dashed lines) and at next-to-leading order
in QCD, ($R^{(1)}$, solid lines), as a function of the 
\MSbar renormalization scale $Q$. The
left panel shows the results for $m_{\tilde t_1} = 120$ GeV and the right
panel for $m_{\tilde t_1} = 225$ GeV, corresponding to approximately
$m_{\eta_{\tilde t}} = 238$ GeV and 447 GeV, respectively.}
\end{figure}
%%%%%%%%%%%%%%%%%%%%%%%%%%%%%%%%%%%%%%%%%%%%%%%%%%%%%%%%%%%%%%%%%%%%%%%%%%%%
In figure \ref{fig:RvsQ}, we compare the predicted ratios $R^{(0)}$ and
$R^{(1)}$ as a function of the choice of \MSbar renormalization scale $Q$,
for two different choices of stop mass $120$ and $225$ GeV, 
corresponding to stoponium masses $m_{\eta_{\tilde t}} = 238$ and
$447$ GeV. The leading-order prediction has a strong scale dependence, which is
not surprising since it is inversely proportional to $\alpha_S^2$. 
The next-to-leading
order prediction for the branching ratio to photons is considerably smaller
for all values of $Q$ considered. For the choice $Q = m_{\tilde t_1}$, the
decrease is roughly 30\%. This may be considered somewhat unfortunate for the
observability of the signal. We also note that the $Q$ dependence of the
predicted branching ratio is much improved by the next-to-leading order
calculation.

%%%%%%%%%%%%%%%%%%%%%%%%%%%%%%%%%%%%%%%%%%%%%%%%%%%%%%%%%%%%%%%%%%%%%%%%%%%%%%%%
\begin{figure}[!tp]
\begin{minipage}[]{0.47\linewidth}
\caption{\label{fig:Rvsm}
The dependence of the ratio of decay widths of stoponium
into $\gamma\gamma+X$ and into hadrons is shown at tree level
($R^{(0)}$, dashed lines) and at next-to-leading order
in QCD, ($R^{(1)}$, solid lines), as a function of the
stoponium mass $m_{\eta_{\tilde t}} \approx 2 m_{\tilde t_1}$. The lower
line in each case is the result for the renormalization scale choice $Q 
=
m_{\tilde t_1}/2$,  and the upper line is for $Q = 2 m_{\tilde t_1}$.}
\end{minipage}
\begin{minipage}[]{0.52\linewidth}
\includegraphics[width=7.8cm,angle=0]{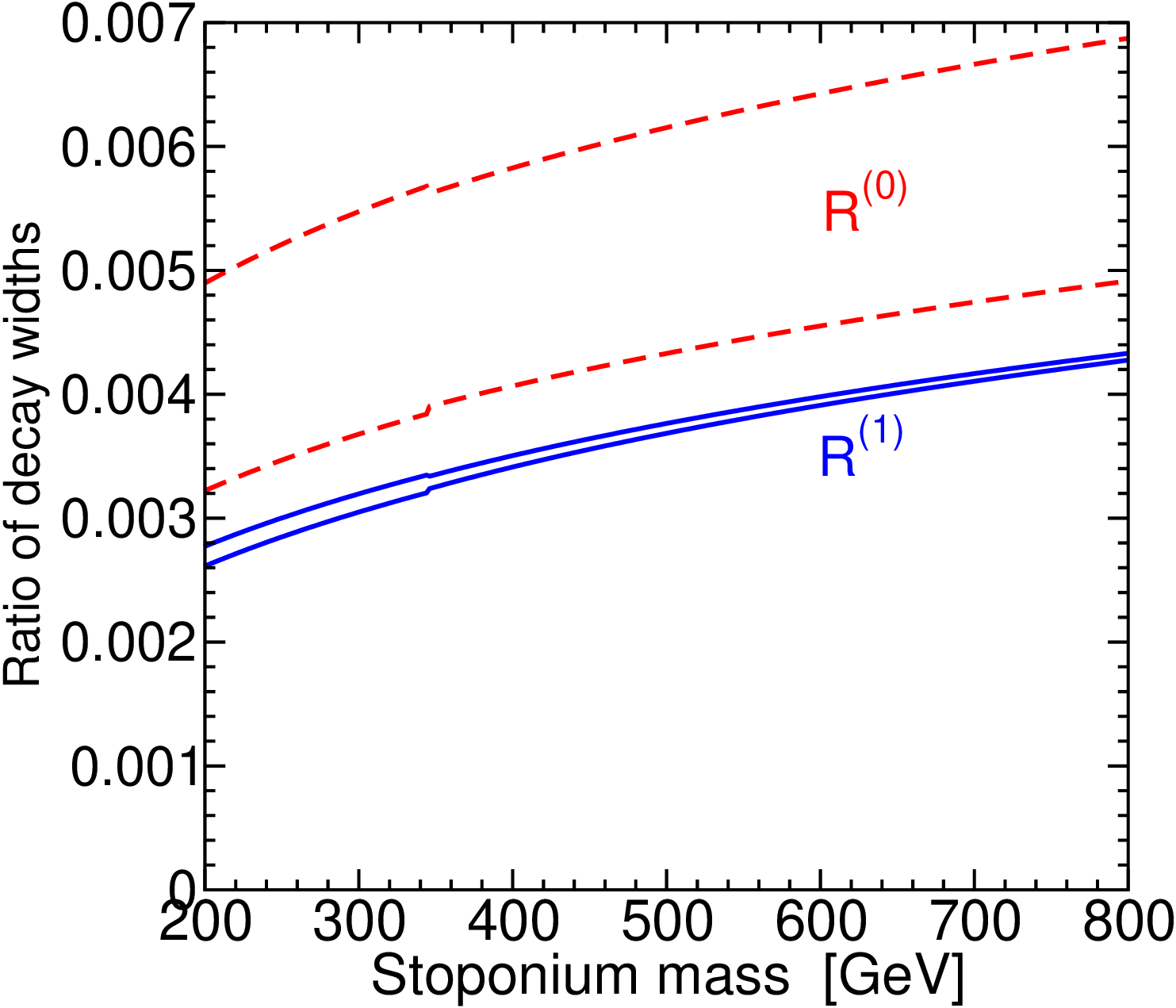}
\end{minipage}
\end{figure}
%%%%%%%%%%%%%%%%%%%%%%%%%%%%%%%%%%%%%%%%%%%%%%%%%%%%%%%%%%%%%%%%%%%%%%%%%%%%%%%%
Figure \ref{fig:Rvsm} shows the ratio of diphoton and hadronic decay widths
as a function of the stoponium mass, computed at leading order and at 
next-to-leading order. The results are shown for two choices 
of the renormalization scale 
$Q = m_{\tilde t_1}/2$ and $Q = 2 m_{\tilde t_1}$.
Again we see that the diphoton branching ratio predicted by the next-to-leading
order calculation is smaller than predicted at leading 
order.\footnote{Small kinks 
are visible in the prediction curves 
in Figure \ref{fig:Rvsm}
at $m_{\eta_{\tilde t}} = 2 m_t$, 
due to our use of the 5-quark (6-quark) effective theory below (above)
that threshold.} Taking into
account the fact that the 
contributions of other final states ($h^0h^0$,
$W^+W^-$, $ZZ$, and $t\overline t$) are quite model-dependent but 
can only decrease the diphoton
branching ratio, we conclude that BR$(\gamma\gamma)$
is not more than about 0.25\% for $m_{\eta_{\tilde t}} = 200$ GeV
and 0.35\% for $m_{\eta_{\tilde t}} = 500$ GeV.

It is somewhat more difficult to make a model-independent statement about the
impact of the QCD radiative corrections for the $h^0h^0$ final state.
However, one can note that the factor $K'$ appearing in 
eq.~(\ref{eq:NLOhhresult})
is generally quite close to unity. Typical numerical values found 
from eq.~(\ref{eq:defKprime}) are
%% version 2: next two lines corrected
$0.006 < K'-1 < 0.021$ for $Q = m_{\tilde t_1}$ and
$0.050 < K'-1 < 0.065$ for $Q = 2 m_{\tilde t_1}$.
The lower range in each case occurs closer to threshold for the
decay. Neglecting this small deviation from unity, one finds 
from eq.~(\ref{eq:gammagammaoverhh})
\beq
\frac{{\rm BR}(\gamma\gamma)}{{\rm BR}(h^0h^0)} \bigg |_{\rm NLO}
=   
\frac{{\rm BR}(\gamma\gamma)}{{\rm BR}(h^0h^0)} \bigg |_{\rm LO}
\Bigl ( 1 - C_F \frac{\alpha_S}{\pi} \Bigl [
\frac{3}{2} \ln (Q^2/4 m_{\tilde t_1}^2) + 4 + 5 \lntwo
\Bigr ] \Bigr ), 
\label{eq:schannelonly}
\eeq
yielding a correction of order ($-23$\%, $-32$\%) 
for $Q/m_{\tilde t_1} = (1, 2)$, respectively.
This correction is comparable to, and has the same sign
as, the ratio of the diphoton to the hadronic branching ratios.
However, we note that the ${\rm BR}(h^0h^0)$ is itself highly model-dependent.
It is generally quite small in the compressed supersymmetry models where
top squarks mediate the annihilation of dark matter in the early universe,
but can be very important in the models with light stops motivated by
baryogenesis.

%%%%%%%%%%%%%%%%%%%%%%%%%%%%%%%%%%%%%%%%%%%%%%%%%%%%%%%%%%%%%%%%%%%%%%%%%%%%%%%%
\section{Outlook\label{sec:outlook}}
\setcounter{footnote}{1}
\setcounter{equation}{0}

In this paper, we have calculated the stoponium decay rates to $gg$, 
$\gamma \gamma$ and $h^0h^0$ at next-to-leading order in QCD, in order to 
obtain the branching ratio to two photons. Our calculation applies to the 
$S$-wave $J^{PC} = 0^{++}$ states, including the ground state that has 
the largest direct production cross-section. We have not included the 
$W^+W^-$ and $Z^0Z^0$ partial widths, which are often the next-largest 
contributions to the total stoponium decay rate, and have neglected the 
remaining decay channels, which have branching ratios of no more than a 
few percent at leading order.\footnote{
This includes the $t\overline t$ final state, which is dominated by 
the $gg$ final state and so has a branching ratio 
of only a few percent even when $m_{\tilde t_1}$ is much larger 
than $m_t$, unless there is a resonance Higgs exchange contribution.}
However, it is worth noting that we can 
approximate the radiative corrections to the decay widths to $W^+W^-$ and 
$Z^0Z^0$, provided that the contributions from t-channel $\tilde b$ and 
$\tilde t$ squark exchange can be neglected. In this approximation, the 
radiative corrections come exclusively from 4-point and $s$-channel 
diagrams similar to H1a through H2b of figure \ref{fig:higgsNLO}, and it 
follows that the ratio of partial widths takes exactly the same form as 
eq.~(\ref{eq:schannelonly}) with the Higgs branching ratio 
${{\rm BR}(h^0h^0)}$ replaced by ${{\rm BR}(W^+W^-)}$ or 
${{\rm BR}(Z^0Z^0)}$ as appropriate.

This simplification applies to both of the model scenarios that motivate 
our study. In order to generate a relic density of dark matter in 
agreement with observation, models with compressed supersymmetry should 
have an LSP between about 170 and 270 GeV and a lightest stop mass 
eigenstate approximately 25 to 100 GeV heavier, which implies a stoponium 
mass very roughly between 400 and 700 GeV \cite{compressed}.  Within this 
range, there is extra suppression in the $W^+W^-$ and $Z^0Z^0$ decay 
channels due to a cancellation between the terms of eq. (A.9) in ref. 
\cite{Martin:2008sv} that come from the 4-point and s-channel Higgs 
exchange diagrams (see also figure 5 of that paper for typical branching 
ratios as a function of stoponium mass in compressed supersymmetry).  
Even with a stoponium mass as high as 600 GeV, the combined $W^+W^-$ and 
$Z^0Z^0$ decay rates represent no more than ten percent of the total 
width.  In any case, the radiative corrections to the model-dependent 
terms arising from t-channel squark exchange are never large enough to 
make a big difference in the total BR($\gamma\gamma$).

In contrast, in models that feature electroweak-scale baryogenesis, the 
combined contribution to the stoponium decay rate from $W^+W^-$ and 
$Z^0Z^0$ can be as large as about 50\% at leading order (see figure 8 in 
ref. \cite{Martin:2008sv}).  However, in this scenario $W^+W^-$ and 
$Z^0Z^0$ decays from t-channel squark exchange are highly suppressed 
because the light stop consists almost entirely of the right-handed gauge 
eigenstate and the mass of the lightest sbottom is much greater than the 
mass of the lightest stop (for leading order amplitudes, see eqs. (A.5) 
through (A.10) of ref. \cite{Martin:2008sv}).  Therefore the ratios of 
the $\gamma \gamma$ partial width to the $W^+W^-$ and $Z^0Z^0$ partial 
widths will take the same form as eq. (\ref{eq:schannelonly}), with 
BR($W^+W^-$) or BR($Z^0Z^0$) in place of BR($h^0h^0$), to a very good 
approximation.

It should be noted \cite{Drees:1993uw} that the $S$-wave excited states 
of stoponium can be produced at the LHC with either direct decays to two 
photons or decays (with emission of soft mesons or photons) to lower 
stoponium states before annihilation. The treatment of direct 
annihilation decays of the excited $S$-wave states is the same as for the 
ground state. These contributions will be essentially indistinguishable 
in the $\gamma\gamma$ invariant mass signal, since the binding energy 
differences between such states will be less than the experimental 
resolution. However, part of the resulting signal will be lost when the 
excited $S$-wave states decay to $P$-wave states that decay directly 
\cite{Drees:1993uw}. Evaluating the contributions from excited states 
will require a more detailed understanding of the stoponium spectroscopy.

Although radiative corrections decrease the branching ratio 
significantly, stoponium annihilation to $\gamma\gamma$ is still a viable 
signal at the LHC. The approximately 30-35\% reduction in the diphoton 
branching ratio found in this paper is likely to be offset by an 
enhancement due to radiative corrections to the cross-section for 
stoponium production in hadron colliders; note that the corresponding 
calculation for toponium production results in a K factor of roughly 1.3 
or more \cite{Kuhn:1992qw}. We plan to report on the corresponding 
results for stoponium in a future paper. In any case, with sufficient 
integrated luminosity, stoponium should be visible, and the diphoton 
decay channel remains a unique opportunity for the direct and precise 
measurement of superpartner masses. Moreover, a measurement of the rate 
for $pp \rightarrow \eta_{\tilde t} \rightarrow \gamma\gamma$, taking 
into account the radiative corrections to both production and decay, will 
provide interesting and useful information about the stoponium system.

\textit{Acknowledgments:} This work was supported in part by National 
Science Foundation grant PHY-0757325.

%%%%%%%%%%%%%%%%%%%%%%%%%%%%%%%%%%%%%%%%%%%%%%%%%%%%%%%%%%%%%%%%%%%%%%%%%%%%%%%%
%%%%%%%%%%%%%%%%%%%%%%%%%%%%%%%%%%%%%%%%%%%%%%%%%%%%%%%%%%%%%%%%%%%%%%%%%%%%%%%%


\begin{thebibliography}{90}
\baselineskip=12.6pt

\bibitem{reviews}
  M.~Drees, R.~Godbole and P.~Roy,
  ``Theory and phenomenology of sparticles: An account of 
four-dimensional N=1
  supersymmetry in high energy physics,''
{\it  World Scientific (2004)};
  H.~Baer and X.~Tata,
  ``Weak scale supersymmetry: From superfields to scattering events,''
{\it  Cambridge University Press (2006)};
  S.P.~Martin,
  ``A supersymmetry primer,''
  [hep-ph/9709356] (version 5, December 2008).
  %%CITATION = HEP-PH/9709356;%%
We use the notations of the last reference.

\bibitem{ATLASTDR}
The ATLAS collaboration,
``ATLAS Detector and physics performance technical design report", Volume 
2.
CERN-LHCC-99-15, ATLAS-TDR-15, May 1999.

\bibitem{CMSTDR}
The CMS collaboration,
``CMS Physics Technical Design Report", Volume 1, Detector Performance
and Software.
CERN-LHCC-2006-001, CMS TDR 8.1, February 2006.

\bibitem{compressed}
S.P.~Martin,
  %``Compressed supersymmetry and natural neutralino dark matter from top
  %squark-mediated annihilation to top quarks,''
  Phys.\ Rev.\  D {\bf 75}, 115005 (2007)
  [hep-ph/0703097],
  %%CITATION = PHRVA,D75,115005;%%
  %``The top squark-mediated annihilation scenario and direct detection of dark
  %matter in compressed supersymmetry,''
  Phys.\ Rev.\  D {\bf 76}, 095005 (2007)
  [hep-ph/0707.2812].
  %%CITATION = PHRVA,D76,095005;%%
  %``Exploring compressed supersymmetry with same-sign top quarks at the Large
  %Hadron Collider,''
  Phys.\ Rev.\  D {\bf 78}, 055019 (2008)
  [hep-ph/0807.2820].
  %%CITATION = PHRVA,D78,055019;%%

\bibitem{WMAP}
  D.N.~Spergel {\it et al.}  [WMAP Collaboration],
  ``Wilkinson Microwave Anisotropy Probe (WMAP) three year results:
  Implications for cosmology,''
  Astrophys.\ J.\ Suppl.\  {\bf 170}, 377 (2007)
  [astro-ph/0603449].
  %%CITATION = ASTRO-PH/0603449;%%

\bibitem{SDSS}
  M.~Tegmark {\it et al.}  [SDSS Collaboration],
  %``Cosmological parameters from SDSS and WMAP,''
  Phys.\ Rev.\  D {\bf 69}, 103501 (2004)
  [astro-ph/0310723].
  %%CITATION = PHRVA,D69,103501;%%

\bibitem{PDG}
W.M.~Yao {\it et al.}  [Particle Data Group],
  ``Review of particle physics,''
  J.\ Phys.\ G {\bf 33}, 1 (2006).
  %%CITATION = JPHGB,G33,1;%%

\bibitem{stopcoannihilationone}
M.E.~Gomez, G.~Lazarides and C.~Pallis,
  %``Supersymmetric cold dark matter with Yukawa unification,''
  Phys.\ Rev.\ D {\bf 61}, 123512 (2000)
  [hep-ph/9907261];
  %%CITATION = HEP-PH 9907261;%%
C.~Boehm, A.~Djouadi and M.~Drees,
  %``Light scalar top quarks and supersymmetric dark matter,''
  Phys.\ Rev.\ D {\bf 62}, 035012 (2000)
  [hep-ph/9911496];
  %%CITATION = HEP-PH 9911496;%%
%
J.R.~Ellis, K.A.~Olive and Y.~Santoso,
  %``Calculations of neutralino stop coannihilation in the CMSSM,''
  Astropart.\ Phys.\  {\bf 18}, 395 (2003)
  [hep-ph/0112113].
  %%CITATION = HEP-PH 0112113;%%

\bibitem{baryoDM}
  C.~Balazs, M.S.~Carena and C.E.M.~Wagner,
%``Dark matter, light stops and electroweak baryogenesis,''
  Phys.\ Rev.\  D {\bf 70}, 015007 (2004)
  [hep-ph/0403224].
  %%CITATION = PHRVA,D70,015007;%%
%
  C.~Balazs, M.S.~Carena, A.~Menon, D.E.~Morrissey and C.E.M.~Wagner,
%``The supersymmetric origin of matter,''
  Phys.\ Rev.\  D {\bf 71}, 075002 (2005)
  [hep-ph/0412264].
  %%CITATION = PHRVA,D71,075002;%%
%  

\bibitem{stopcoannihilationthree}
G.~Belanger, F.~Boudjema, S.~Kraml, A.~Pukhov and A.~Semenov,
  %``Relic density of neutralino dark matter in the MSSM with CP violation,''
  Phys.\ Rev.\  D {\bf 73}, 115007 (2006)
  [hep-ph/0604150].
  %%CITATION = PHRVA,D73,115007;%%

\bibitem{baryo}
J.R.~Espinosa, M.~Quiros and F.~Zwirner,
  %``On the electroweak phase transition in the minimal supersymmetric Standard
  %Model,''
  Phys.\ Lett.\  B {\bf 307}, 106 (1993)
  [hep-ph/9303317],
  %%CITATION = PHLTA,B307,106;%%
%
M.S.~Carena, M.~Quiros and C.E.M.~Wagner,
  %``Opening the Window for Electroweak Baryogenesis,''
  Phys.\ Lett.\  B {\bf 380}, 81 (1996)
  [hep-ph/9603420],
  %%CITATION = PHLTA,B380,81;%%
%
%M.S.~Carena, M.~Quiros and C.E.M.~Wagner,
  %``Electroweak baryogenesis and Higgs and stop searches at LEP and the
  %Tevatron,''
  Nucl.\ Phys.\  B {\bf 524}, 3 (1998)  
  [hep-ph/9710401],
  %%CITATION = NUPHA,B524,3;%%
%
J.R.~Espinosa,
  %``Dominant Two-Loop Corrections to the MSSM Finite Temperature Effective
  %Potential,''
  Nucl.\ Phys.\  B {\bf 475}, 273 (1996) 
  [hep-ph/9604320],
  %%CITATION = NUPHA,B475,273;%%
%
D.~Bodeker, P.~John, M.~Laine and M.~G.~Schmidt,
  %``The 2-loop MSSM finite temperature effective potential with stop
  %condensation,''
  Nucl.\ Phys.\  B {\bf 497}, 387 (1997)
  [hep-ph/9612364],
  %%CITATION = NUPHA,B497,387;%%
%
M.S.~Carena, M.~Quiros, A.~Riotto, I.~Vilja and C.E.M.~Wagner,
  %``Electroweak baryogenesis and low energy supersymmetry,''
  Nucl.\ Phys.\  B {\bf 503}, 387 (1997)
  [hep-ph/9702409],
  %%CITATION = NUPHA,B503,387;%%
%
J.M.~Cline, M.~Joyce and K.~Kainulainen,
  %``Supersymmetric electroweak baryogenesis in the WKB approximation,''
  Phys.\ Lett.\  B {\bf 417}, 79 (1998)
  [Erratum-ibid.\  B {\bf 448}, 321 (1999)]
  [hep-ph/9708393],
  %%CITATION = PHLTA,B417,79;%%
%J.~M.~Cline, M.~Joyce and K.~Kainulainen,
  %``Supersymmetric electroweak baryogenesis,''
  JHEP {\bf 0007}, 018 (2000)   
  [hep-ph/0006119],
  %%CITATION = JHEPA,0007,018;%%
J.M.~Cline and G.D.~Moore,
  %``Supersymmetric electroweak phase transition: Baryogenesis versus
  %experimental constraints,''
  Phys.\ Rev.\ Lett.\  {\bf 81}, 3315 (1998)
  [hep-ph/9806354],
  %%CITATION = PRLTA,81,3315;%%
%
M.S.~Carena, M.~Quiros, M.~Seco and C.E.M.~Wagner,
%``Improved results in supersymmetric electroweak baryogenesis,''
  Nucl.\ Phys.\  B {\bf 650}, 24 (2003)
  [hep-ph/0208043].
  %%CITATION = NUPHA,B650,24;%%
  
\bibitem{baryonew} 
M.~Carena, G.~Nardini, M.~Quiros and C.E.M.~Wagner,
  %``The Effective Theory of the Light Stop Scenario,''
  JHEP {\bf 0810}, 062 (2008)  [hep-ph/0806.4297].  
  %%CITATION = JHEPA,0810,062;%% 
M.~Carena, A.~Freitas and C.E.M.~Wagner,  
  %``Light Stop Searches at the LHC in Events with One Hard Photon or Jet and  
  %Missing Energy,''  
  JHEP {\bf 0810}, 109 (2008)  [hep-ph/0808.2298].  
  %%CITATION = JHEPA,0810,109;%%
M.~Carena, G.~Nardini, M.~Quiros and C.E.M.~Wagner,  
  ``The Baryogenesis Window in the MSSM,''  [hep-ph/0809.3760].  
  %%CITATION = ARXIV:0809.3760;%%

%% version 2: reference added
\bibitem{DiazCruz:2007fc}
 J.L.~Diaz-Cruz, J.R.~Ellis, K.A.~Olive and Y.~Santoso,
 %``On the feasibility of a stop NLSP in gravitino dark matterscenarios,'' 
 JHEP {\bf 0705}, 003 (2007)
 [hep-ph/0701229].
 %%CITATION = JHEPA,0705,003;%%

\bibitem{Hikasa:1987db}
  K.-i.~Hikasa and M.~Kobayashi,
  %``LIGHT SCALAR TOP AT e+ e- COLLIDERS,''
  Phys.\ Rev.\  D {\bf 36}, 724 (1987).
  %%CITATION = PHRVA,D36,724;%%

\bibitem{Boehm:1999tr}
  C.~Boehm, A.~Djouadi and Y.~Mambrini,
  %``Decays of the lightest top squark,''
  Phys.\ Rev.\  D {\bf 61}, 095006 (2000)
  [hep-ph/9907428].
  %%CITATION = PHRVA,D61,095006;%%

\bibitem{SDECAY}
  M.~Muhlleitner, A.~Djouadi and Y.~Mambrini,
  %``SDECAY: A Fortran code for the decays of the supersymmetric particles  in
  %the MSSM,''
  Comput.\ Phys.\ Commun.\  {\bf 168}, 46 (2005)
  [hep-ph/0311167].
  %%CITATION = CPHCB,168,46;%%

\bibitem{Das:2001kd}
  S.P.~Das, A.~Datta and M.~Guchait,
  %``Four body decay of the stop squark at the upgraded Tevatron,''
  Phys.\ Rev.\  D {\bf 65}, 095006 (2002)
  [hep-ph/0112182].
  %%CITATION = PHRVA,D65,095006;%%

\bibitem{Hiller:2008wp}
  G.~Hiller and Y.~Nir,
  %``Measuring Flavor Mixing with Minimal Flavor Violation at the LHC,''
  JHEP {\bf 0803}, 046 (2008)
  [hep-ph/0802.0916].
  %%CITATION = JHEPA,0803,046;%%

\bibitem{Hagiwara:1990sq}
  K.~Hagiwara, K.~Kato, A.D.~Martin and C.K.~Ng,
  %``Properties Of Heavy Quarkonia And Related States,''
  Nucl.\ Phys.\  B {\bf 344}, 1 (1990).
  %%CITATION = NUPHA,B344,1;%%  

\bibitem{Drees:1993yr}
  M.~Drees and M.M.~Nojiri,
  %``A New Signal For Scalar Top Bound State Production,''
  Phys.\ Rev.\ Lett.\  {\bf 72}, 2324 (1994)
  [hep-ph/9310209].
  %%CITATION = PRLTA,72,2324;%%
  
\bibitem{Drees:1993uw}
  M.~Drees and M.M.~Nojiri,
  %``Production and decay of scalar stoponium bound states,''
  Phys.\ Rev.\  D {\bf 49}, 4595 (1994)
  [hep-ph/9312213].
  %%CITATION = PHRVA,D49,4595;%%

\bibitem{Nappi:1981ft}
  C.R.~Nappi,
  %``Spin 0 Quarks In E+ E- Annihilation,''
  Phys.\ Rev.\  D {\bf 25}, 84 (1982).
  %%CITATION = PHRVA,D25,84;%%

\bibitem{Moxhay:1985bg}
  P.~Moxhay and R.W.~Robinett,
  %``Searching For Scalar Quarkonium At Proton - Anti-Proton Colliders,''
  Phys.\ Rev.\  D {\bf 32}, 300 (1985).
  %%CITATION = PHRVA,D32,300;%%

\bibitem{Herrero:1987df}
  M.J.~Herrero, A.~Mendez and T.G.~Rizzo,
  %``PRODUCTION OF HEAVY SQUARKONIUM AT HIGH-ENERGY p p COLLIDERS,''
  Phys.\ Lett.\  B {\bf 200}, 205 (1988).
  %%CITATION = PHLTA,B200,205;%%

\bibitem{Barger:1988sp}
  V.D.~Barger and W.Y.~Keung,
  %``STOPONIUM DECAYS TO HIGGS BOSONS,''
  Phys.\ Lett.\  B {\bf 211}, 355 (1988).
  %%CITATION = PHLTA,B211,355;%%

\bibitem{Inazawa:1993qk}
  H.~Inazawa and T.~Morii,
  %``T Anti-T Bound State Production At Multi - Tev Hadron Colliders,''
  Phys.\ Rev.\ Lett.\  {\bf 70}, 2992 (1993).
  %%CITATION = PRLTA,70,2992;%%

\bibitem{Gorbunov:2000tr}
  D.S.~Gorbunov and V.A.~Ilyin,
  %``Stoponium search at photon linear collider,''
  JHEP {\bf 0011}, 011 (2000)
  [hep-ph/0004092].
  %%CITATION = JHEPA,0011,011;%%

\bibitem{Fabiano:2001cw}
  N.~Fabiano,
  %``Estimates of threshold cross section for stoponium production at e+ e-
  %colliders,''
  Eur.\ Phys.\ J.\  C {\bf 19}, 547 (2001)
  [hep-ph/0103006].
  %%CITATION = EPHJA,C19,547;%%

%\cite{Martin:2008sv}
\bibitem{Martin:2008sv}
  S.P.~Martin,
  %``Diphoton decays of stoponium at the Large Hadron Collider,''
  Phys.\ Rev.\  D {\bf 77}, 075002 (2008)
  [hep-ph/0801.0237].
  %%CITATION = PHRVA,D77,075002;%%

%\cite{Hagiwara:1980nv}
\bibitem{Hagiwara:1980nv}
  K.~Hagiwara, C.B.~Kim and T.~Yoshino,
  %``Hadronic Decay Rate Of Ground State Paraquarkonia In Quantum
  %Chromodynamics,''
  Nucl.\ Phys.\  B {\bf 177}, 461 (1981).
  %%CITATION = NUPHA,B177,461;%%

\bibitem{Barbieri:1979be}
  R.~Barbieri, E.~d'Emilio, G.~Curci and E.~Remiddi,
  %``Strong Radiative Corrections To Annihilations Of Quarkonia In QCD,''
  Nucl.\ Phys.\  B {\bf 154}, 535 (1979).
  %%CITATION = NUPHA,B154,535;%%
    
\bibitem{Novikov}
V.A.~Novikov, L.B.~Okun, M.A.~Shifman, A.I.~Vainshtein,
M.B.~Voloshin and V.I.~Zakharov,
 %``Charmonium And Gluons: Basic Experimental Facts And Theoretical
 %Introduction,''
 Phys.\ Rept.\  {\bf 41}, 1 (1978).
 %%CITATION = PRPLC,41,1;%%

\bibitem{Beenakker:1988bq}
  W.~Beenakker, H.~Kuijf, W.L.~van Neerven and J.~Smith,
  %``QCD Corrections to Heavy Quark Production in p anti-p Collisions,''
  Phys.\ Rev.\  D {\bf 40}, 54 (1989).
  %%CITATION = PHRVA,D40,54;%%

%\cite{Ellis:2007qk}
\bibitem{Ellis:2007qk}
  R.K.~Ellis and G.~Zanderighi,
  %``Scalar one-loop integrals for QCD,''
  JHEP {\bf 0802}, 002 (2008)
  [hep-ph/0712.1851].
  %%CITATION = JHEPA,0802,002;%%

\bibitem{SPMcouplingsnotation}
  S.P.~Martin,
  %``Two-loop effective potential for the minimal supersymmetric standard
  %model,''
  Phys.\ Rev.\  D {\bf 66}, 096001 (2002)
  [hep-ph/0206136],
  %%CITATION = PHRVA,D66,096001;%%
%
%\bibitem{Martin:2004kr}
 % S.P.~Martin,
  %``Strong and Yukawa two-loop contributions to Higgs scalar boson
  %self-energies and pole masses in supersymmetry,''
  Phys.\ Rev.\  D {\bf 71}, 016012 (2005)
  [hep-ph/0405022].
  %%CITATION = PHRVA,D71,016012;%%

\bibitem{tHooft:1978xw} 
G.~'t Hooft and M.J.G.~Veltman, 
  %``Scalar One Loop Integrals,'' 
  Nucl.\ Phys.\ B {\bf 153}, 365 (1979).  
  %%CITATION = NUPHA,B153,365;%%

\bibitem{ref:coulsing}
I.~Harris and L.M.~Brown,
  %``Radiative Corrections to Pair Annihilation,''
  Phys.\ Rev.\  {\bf 105}, 1656 (1957).
  %%CITATION = PHRVA,105,1656;%%
G.T.~Bodwin, E.~Braaten and G.P.~Lepage,
  %``Rigorous QCD analysis of inclusive annihilation and production of 
heavy
  %quarkonium,''
  Phys.\ Rev.\  D {\bf 51}, 1125 (1995)
  [Erratum-ibid.\  D {\bf 55}, 5853 (1997)]
  [hep-ph/9407339].
  %%CITATION = PHRVA,D51,1125;%%
A.~Petrelli, M.~Cacciari, M.~Greco, F.~Maltoni and M.L.~Mangano,
  %``NLO production and decay of quarkonium,''
  Nucl.\ Phys.\  B {\bf 514}, 245 (1998)
  [hep-ph/9707223].
  %%CITATION = NUPHA,B514,245;%%

%\cite{Srednicki:2007qs}
\bibitem{Srednicki:2007qs}
  See, for example,
  M.~Srednicki,
  ``Quantum field theory,''
%\href{http://www.slac.stanford.edu/spires/find/hep/www?irn=7209290}{SPIRES entry}
{\it  Cambridge, UK: Univ. Pr. (2007)}, eq. (78.36);
D.R.T.~Jones,
  %``Two Loop Diagrams In Yang-Mills Theory,''
  Nucl.\ Phys.\  B {\bf 75}, 531 (1974).
  %%CITATION = NUPHA,B75,531;%%

\bibitem{Kuhn:1992qw}
  J.~H.~Kuhn and E.~Mirkes,
  %``QCD Corrections To Toponium Production At Hadron Colliders,''
  Phys.\ Rev.\  D {\bf 48}, 179 (1993)
  [arXiv:hep-ph/9301204].
  %%CITATION = PHRVA,D48,179;%%

\end{thebibliography}
\end{document}